\documentclass[twocolumn]{aastex701} 

\usepackage{graphicx}	
\usepackage{amsmath}	
\usepackage{changepage}
\usepackage{comment} 
\usepackage{fancyhdr}
\usepackage{booktabs}
\usepackage{gensymb}
\usepackage{tabularx}
\usepackage{ragged2e}
\usepackage{float}
\usepackage{hyperref}
\usepackage{wrapfig}
\usepackage{sidecap}
\usepackage{natbib}

\newcommand{\OI}{\mbox{O\,{\sc i}}}
\newcommand{\HI}{\mbox{H\,{\sc i}}}

\newcommand{\SiII}{\mbox{Si\,{\sc ii}}}
\newcommand{\SiIII}{\mbox{Si\,{\sc iii}}}
\newcommand{\SiIV}{\mbox{Si\,{\sc iv}}}
\newcommand{\CII}{\mbox{C\,{\sc ii}}}

\newcommand{\CIV}{\mbox{C\,{\sc iv}}}
\newcommand{\msun}{\,$\rm M_{\odot}$}
\newcommand{\msyr}{\,$\rm M_{\odot}$\,yr$^{-1}$}
\newcommand{\sqcm}{\,cm$^{-2}$}  
\newcommand{\rvir}{\,$R_{\rm 200}$}     
\newcommand{\kms}{\,km\,s$^{-1}$}

\newcommand{\jone}{SDSS~J100035.48+052428.5} 
\newcommand{\jshort}{J1000+0524}
\newcommand{\jtshort}{J0959+0503}
\newcommand{\tm}{\tablenotemark} 
\newcommand{\tn}{\tablenotetext}

\defcitealias{zheng2024}{Z24}

\begin{document}

\title{Low Metallicity Gas on the Outskirts of the Local Group: the Circumgalactic Medium of Sextans B\footnote{Based on observations made with the NASA/ESA Hubble Space Telescope, obtained at the Space Telescope Science Institute, which is operated by the Association of Universities for Research in Astronomy, Inc., under NASA contract NAS5-26555. These observations are associated with program 17209.}} 

\author[0000-0003-0724-4115]{Andrew J. Fox}
\affiliation{AURA for ESA, Space Telescope Science Institute, 3700 San Martin Drive, Baltimore, MD 21218}
\email{afox@stsci.edu}

\author[0000-0002-4157-5164]{Sapna Mishra}
\affiliation{Space Telescope Science Institute, 3700 San Martin Drive, Baltimore, MD 21218, USA}
\email{smishra@stsci.edu}

\author[0000-0003-4237-3553]{Frances H. Cashman}
\affiliation{Department of Physics, Presbyterian College, Clinton, SC 29325, USA}
\email{fcashman@presby.edu}

\author[0000-0003-3681-0016]{David M. French}
\affiliation{Space Telescope Science Institute, 3700 San Martin Drive, Baltimore, MD 21218, USA}
\email{dfrench@stsci.edu}

\author[0000-0002-1188-1435]{Philipp Richter}
\affiliation{Institut f{\"u}r Physik und Astronomie, Universit{\"a}t Potsdam, Karl-Liebknecht-Str. 24/25, 14476 Golm, Germany}
\email{prichter@astro.physik.uni-potsdam.de}

\author[0000-0002-3120-7173]{Rongmon Bordoloi}
\affiliation{Department of Physics, North Carolina State University, Raleigh, NC 27695, USA}
\email{rbordol@ncsu.edu}

\author[0000-0001-9158-0829]{Nicolas Lehner}
\affiliation{Department of Physics and Astronomy, University of Notre Dame, Notre Dame, IN 46556, USA}
\email{nlehner@nd.edu}

\author[0000-0002-7982-412X]{Jason Tumlinson}
\affiliation{Space Telescope Science Institute, 3700 San Martin Drive, Baltimore, MD 21218, USA}
\affiliation{Department of Physics \& Astronomy, Johns Hopkins University, 3400 N. Charles Street, Baltimore, MD 21218, USA}
\email{tumlinson@stsci.edu}


\author[0000-0002-2724-8298]{Sanchayeeta Borthakur}
\affiliation{School of Earth \& Space Exploration, Arizona State University, 781 Terrace Mall, Tempe, AZ 85287, USA}
\email{sborthak@asu.edu}

\correspondingauthor{Andrew Fox}

\begin{abstract}
We present a UV absorption-line analysis of the circumgalactic medium (CGM) of Sextans B, 
a dwarf irregular galaxy at 1.3 Mpc distance on the outer frontier of the Local Group. Using 
HST/COS spectroscopy of two AGN sightlines passing through the Sextans B CGM at small 
impact parameters of 4 kpc and 8 kpc ($\approx$0.04 to 0.08\rvir), we detect the 
CGM in \SiII, \SiIII, \SiIV, and \CII\ absorption. 
All four ions show a column-density profile that declines with radius.
The profiles fall below the average CGM profiles of other nearby dwarfs 
(by $\approx$0.3--0.6\,dex, depending on ion),
likely due to the low halo mass and low metallicity of Sextans B. 
Using {\it Cloudy} photoionization models and interferometric measurements 
of the \HI\ column density, we find gas-phase silicon and carbon abundances 
in the Sextans B CGM of [Si/H]=$-1.7\pm0.2$ and [C/H]=$-2.1\pm0.2$,
among the lowest gas-phase abundances anywhere in the Local Group.
We calculate a cool CGM gas mass within 8 kpc of 
$\approx4\times10^{7}$\msun,
comparable to the \HI\ mass and the stellar mass of Sextans B.
\end{abstract}

\section{Introduction} \label{sec:intro}

Extending over 1 Mpc away from us, the Local Group (LG) provides a unique 
testbed for galaxy evolution. 
The LG includes about 40 Irregular galaxies \citep[Irrs;][]{mcc2012}, 
ranging in mass from the LMC ($M_*\sim10^{10}$\msun) down to dwarf irregulars 
with $M_*\sim10^{7-8}$\msun. 
Dwarf irregulars tend to have
high star formation rates, high gas masses, and complex morphologies, but the 
reasons for these unusual properties are not fully understood.
Important clues to understanding these properties
lie in their circumgalactic medium (CGM), the extended baryon reservoir that acts
as a source for inflows and a sink for outflows
\citep{putman2012, tumlinson2017}.

Several recent studies have targeted the CGM of individual dwarf galaxies in the LG, 
including WLM \citep{zheng2019}, IC1613 \citep{zheng2020}, and 
Sextans A \citep{qu2022}. More recently, a large sample of 45 nearby 
dwarf-galaxy CGMs was recently presented in \citet[][hereafter Z24]{zheng2024},
including many LG galaxies; they found that the cool CGM harbors $\sim$10\% of the 
metals ever produced in these galaxies \citep[see also][]{bordoloi2014, johnson2017}.

The dwarf irregular galaxy Sextans B (also known as DDO 70 and UGC 5373) lies  
at a distance of 1.3 Mpc \citep{vdb2000} in the LG anti-center direction
at galactic coordinates $l$=233.2\degree, $b$=43.8\degree.
It is part of the NGC\,3109 sub-group  
with NGC\,3109, Sextans A, and the Antlia Dwarf 
\citep{vdb1999, tully2006, bellazzini2013}.
This sub-group exists on the outer fringes of the LG,
far from the MW and M31 systems that dominate the LG.
Sextans B therefore exists in a quiescent and relatively isolated environment 
where the influence of galaxy interactions 
on the CGM is expected to be small \citep[at the current time; 
the NGC\,3109 group is thought to 
have passed close to the Milky Way $\sim$7 Gyr ago, so the interactions may
have been stronger in the past;][]{shaya2013}.
Its halo provides a case study of the CGM of isolated irregular galaxies. 

Sextans B has a stellar mass of
$4.0-5.2\times10^7$\msun \citep[Z24;][]{mcc2012},
an \HI\ mass of $4.1\times10^7$\msun\ \citep{hunter2012},
a stellar metallicity between [Fe/H]=$-$1.9 and $-$1.6 
\citep{vdb2000, bellazzini2014}, and an oxygen abundance in \ion{H}{2} regions 
ranging from [O/H]=$-$1.16 to $-$0.85 \citep{kniazev2005}. 
It has a current FUV star formation rate (SFR) of $3.6\times10^{-3}$\msyr\ 
\citep{hunter2010} and a lifetime-averaged SFR of
$1.18_{-0.95}^{+0.96}\times10^{-3}$\msyr\ \citep{weisz2011}. 
Much of the current star formation is concentrated into \HI\ supershells
\citep{gerasimov2024}.
Sextans B has a dynamical mass of $\sim1-4\times10^9$\msun\ \citep{bellazzini2014}
and a total halo mass of $3.3\times10^{10}$\msun\ using the 
stellar-mass-to-halo-mass relation of \citet{munshi2021}, giving 
a halo radius $R_{\rm 200}$ of 100 kpc\footnote{We use $R_{\rm 200}$ 
(the radius of the region in which the mass density is 200 times the 
critical density) instead of the virial radius $R_{\rm vir}$ to measure the halo size, 
for consistency with earlier work \citepalias{zheng2024}. $R_{\rm 200}$ is slightly less 
than $R_{\rm vir}$.} \citepalias{zheng2024}. 

In this paper we present HST/COS spectroscopy of two AGN sightlines passing 
through the Sextans B CGM at very low impact parameters of
4.1\,kpc and 8.0 kpc ($\approx$0.04\rvir\ and 0.08\rvir). 
This allows us to probe the inner CGM of a dwarf 
irregular galaxy in an isolated environment where galaxy interactions are minimal, 
and to determine the radial profile of its low-ion and high-ion CGM.
Furthermore, only four LG galaxies (not counting the Milky Way) have published CGM 
measurements from \emph{multiple} sightlines, and therefore observationally-constrained 
radial profiles: the LMC \citep{krishnarao2022, mishra2024}, 
M31 \citep{lehner2015, lehner2020, lehner2025, richter2017}, IC1613 \citep{zheng2020}, 
and Sextans A \citep{qu2022}. Our results on Sextans B expand this number to five.

\section{Observations and Data Handling} \label{sec:obs}

\begin{figure}[!ht]
 \includegraphics[width=0.45\textwidth]{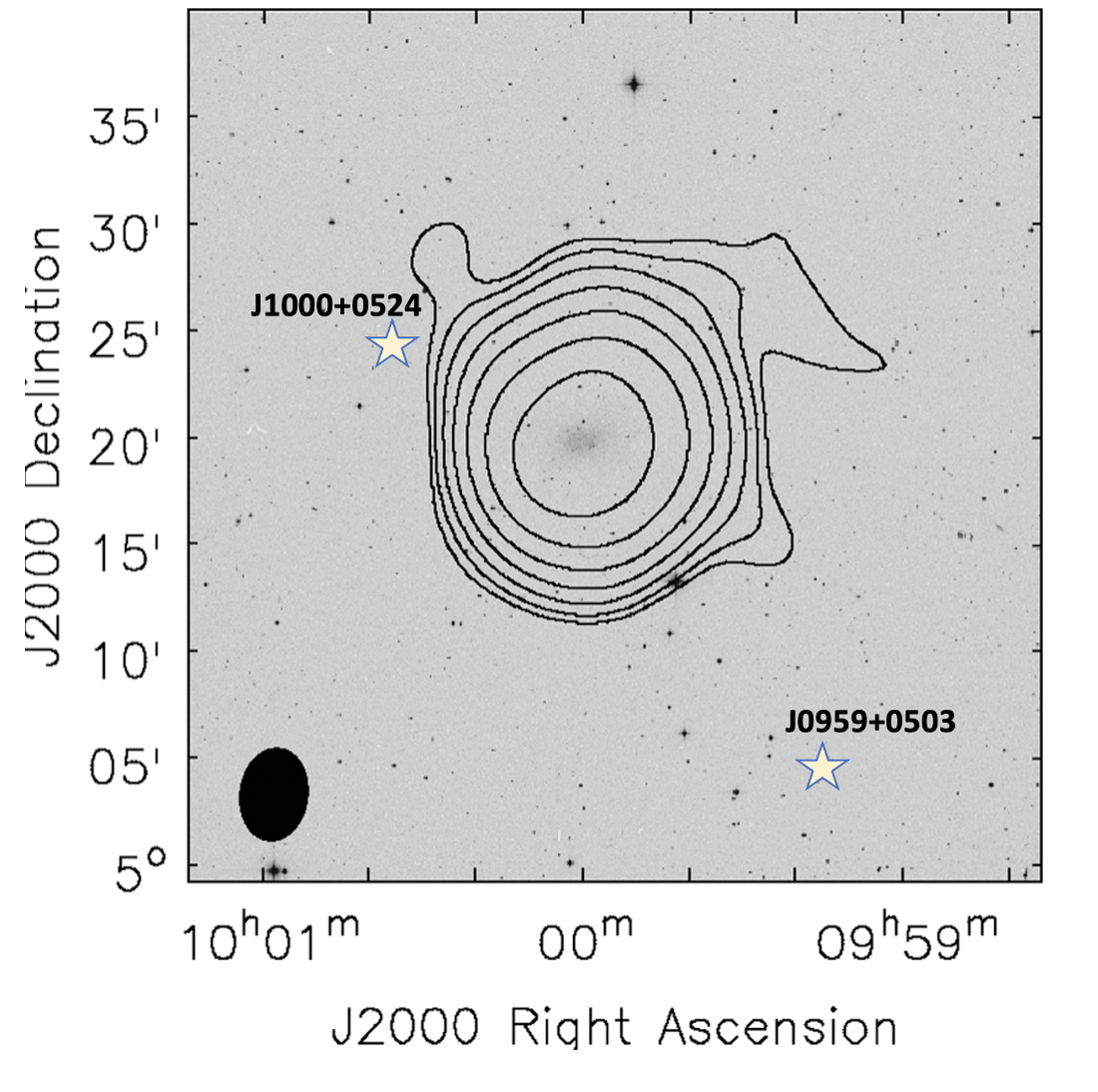}
\caption{Location of the two HST/COS AGN sightlines (stars) relative to the 
 \HI\ 21 cm emission in Sextans B (contours) and a DSS image of the galaxy 
 showing the stellar component \citep[figure adapted from][]{namumba2018}.
 The \HI\ data is from KAT-7 
 with an effective spatial resolution (FWHM) of 4.4$\times$3.2\arcmin.  
 The \HI\ contours are at 5.4, 10.8, 21.6, 43.2, 86.4, 172.8, and 345.6 $\times10^{18}$\sqcm.} 
 \label{fig:map} 
\end{figure}

\begin{figure*}[!ht]
 \includegraphics[width=0.95\textwidth]{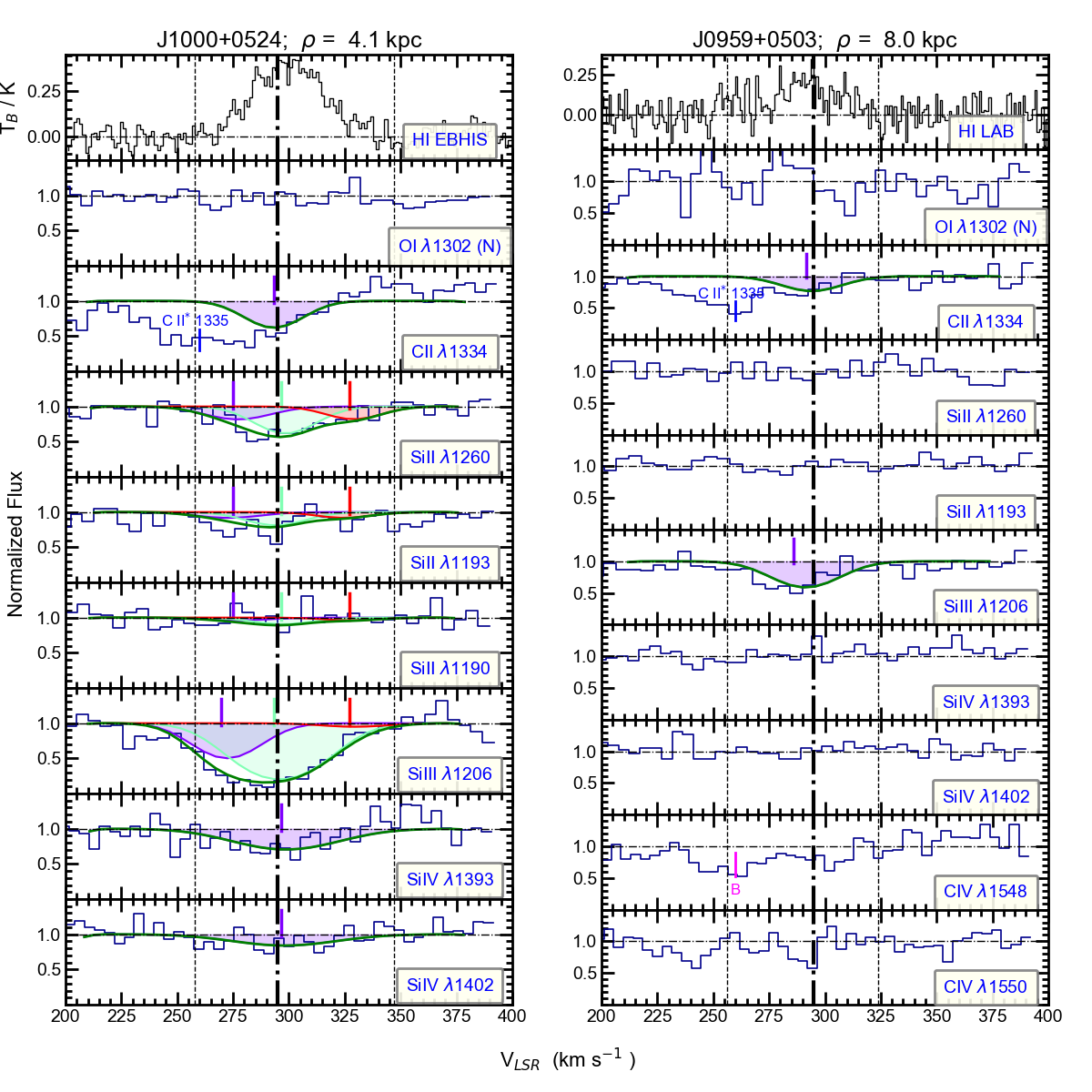}
 \caption{Normalized absorption-line profiles of UV metal lines at the Sextans B velocity 
 in the HST/COS spectra of two AGN, \jshort\ (left, $\rho$=4.1\,kpc) and \jtshort\ 
 (right; $\rho$=8.0\,kpc). The top panel shows the \HI\ 21 cm emission profile from the 
 EBHIS survey \citep{winkel2016} or LAB survey \citep{kalberla2005}. 
 The COS data are binned by two pixels.
 Individual Voigt components are shaded in distinct colors with their 
 central velocities marked with short vertical lines. 
 The thick green line shows the overall model, 
 the thick vertical dot-dashed line shows the systemic velocity of Sextans B (295\kms), 
 and the thin vertical dashed lines show the velocity integration range.
 The label (N) for \OI\ 1302 refers to night-only reduction, and 'B' indicates a blend.
 A third sightline (PG1001+054 at $\rho$=27 kpc) gives no Sextans B detections 
 in any of the UV lines \citepalias{zheng2024} and is not shown.}
 \label{fig:stack_ebhis} 
\end{figure*}

\begin{figure*}[!ht]
 \includegraphics[width=0.95\textwidth]{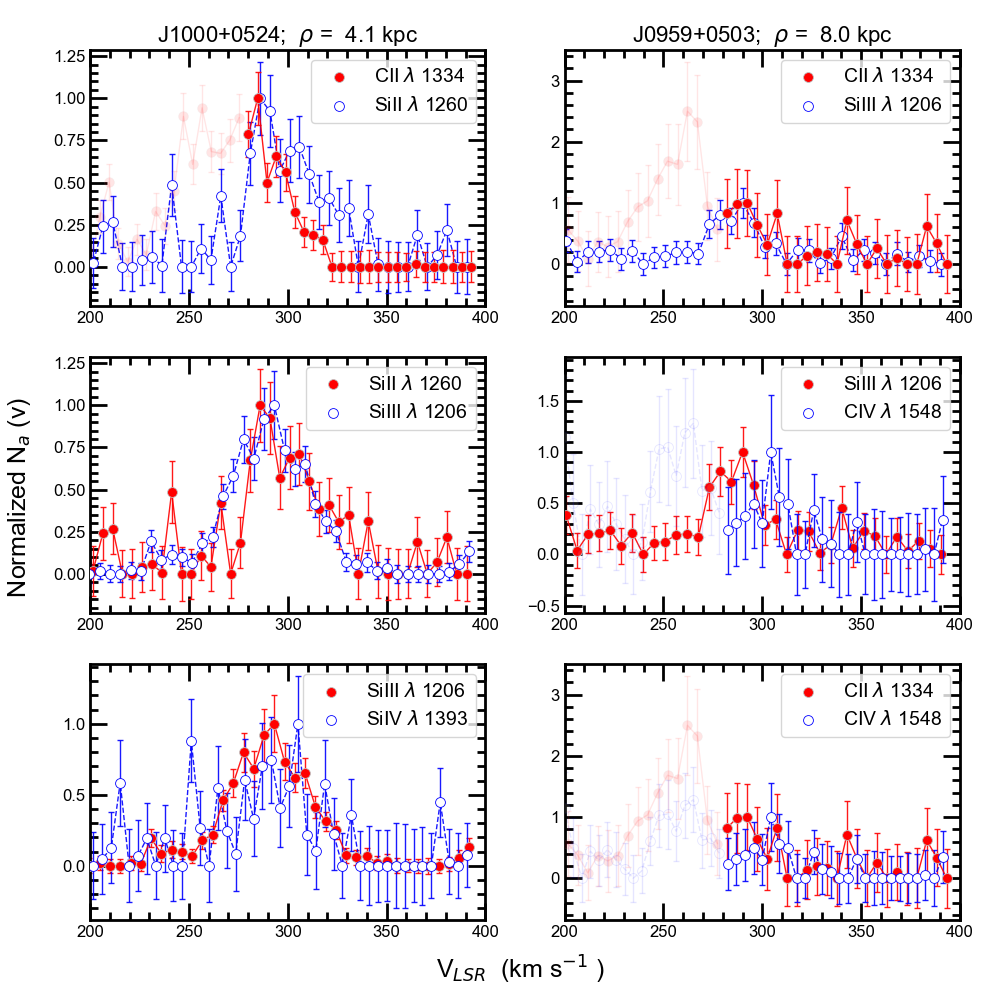}
 \caption{Normalized apparent column density profiles of UV metal absorption in the 
 Sextans B CGM. The left column shows the \jshort\ sightline and the right column shows 
 \jtshort. Different pairs 
 of ions are shown in each panel, allowing us to assess which ions are co-spatial in the 
 Sextans B CGM. Each profile is normalized to its maximum value. Contaminated velocity regions 
 (for \CII\ 1334 and \CIV\ 1548) are shown in pale color and can be ignored.} 
 \label{fig:acd} 
\end{figure*}

\subsection{Target Selection} \label{subsec:target}

We obtained HST/COS FUV spectroscopy of the AGN \jone\ 
(hereafter \jshort; $z_{\rm em}$=0.07858) under HST 
Program ID 17209 (PI: A. Fox). We targeted \jshort\ because it lies 
very close to Sextans B at an impact parameter $\rho$=4.1 kpc, corresponding to only 
0.04\rvir. This makes it the smallest-impact parameter of any AGN sightline 
passing close to a Local Group dwarf galaxy, based on the sample of 54 CGM 
sightlines in \citetalias{zheng2024}.
Very few UV-bright AGN sightlines are found at such low impact parameters,
at any redshift \citep[see examples in][]{stocke2013, qu2019}.

Three other AGN with archival COS FUV spectra lie in the vicinity of Sextans B.
In order of increasing impact parameter, they are
SDSS J095915.60+050355.0 (\jtshort\ for short; $\rho$=8.4 kpc; $z_{\rm em}$=0.16177),
PG 1001+054 ($\rho$=27 kpc; $z_{\rm em}$=0.16012) and
SDSS J100535.24+013445.7 ($\rho$=100 kpc; $z_{\rm em}$=1.077).
The PG 1001+054 and SDSS J100535.24+013445.7 spectra were published in \citetalias{zheng2024} 
\citep[and by][in the case of PG1001+054]{qu2022} 
and show no Sextans B CGM detections in any UV metal absorption lines,
so we do not include them in our analysis.
In contrast, the \jtshort\ spectra contain CGM detections, 
so we downloaded them from the MAST archive and analyzed in an 
identical manner to the \jshort\ observations.
The location of the \jshort\ and \jtshort\ sightlines is shown 
on a high-resolution \HI\ 
map of Sextans B in Figure~\ref{fig:map}. These \HI\ data were taken with the 
Karoo Array Telescope \citep[KAT-7;][]{foley2016, namumba2018}.

\subsection{HST/COS Observations} \label{subsec:cos}

The COS FUV observations of \jshort\ were taken with the G130M/1291 setting 
(covering 1137--1432\AA) with two FP-POS positions (3 and 4), for a total of 
36,036 s of exposure time across four visits (there were initially two 
four-orbit visits, but guiding failures led to two repeat observations being 
necessary). The data were taken at COS FUV Lifetime Position LP5, which for 
the G130M/1291 setting has a spectral resolution of 20\kms\ (FWHM). 

To reduce and coadd the spectra, we followed the procedures described in 
\citet{krishnarao2022} and \citet{mishra2024a}.
In brief, we used the standard reductions ({\it x1d} files) of the HST/COS spectra 
provided by the {\bf calcos} data reduction pipeline, and
coadded the reduced spectra across visits using the STScI HASP software \citep{debes2024}
to produce a final science spectrum. Finally we binned the spectra by two pixels, 
and defined both global and local continua around each line of interest.
We also conducted a night-only reduction 
to remove the effects of geocoronal emission from the \OI\ 1302 line.
The final \jshort\ (\jtshort) spectrum has a S/N ratio of 14 (20) per 6-pixel 
resolution element at 1200\AA.

We used the Voigt-profile fitting software \texttt{VPFIT} (v12.3, \citealt{Carswell2014}) 
to model the UV metal absorption profiles from Sextans B.
\texttt{VPFIT} determines the line centroid, Doppler parameter ($b$), and 
column density ($N$) for each absorption component using a chi-squared 
minimization process. The component structure was determined by visual 
inspection of the line profiles.
We fit the following metal lines: \CII\ $\lambda$1334, 
\SiII\ $\lambda\lambda$1260,1193,1190 
($\lambda$1304 is not detected), \SiIII\ $\lambda$1206, and 
\SiIV\ $\lambda\lambda$1393,1402. 
We found that a similar component structure applied to all the detected lines. 
For the \jtshort\ direction, which has G160M data as well as G130M, 
we inspected the \CIV\ $\lambda\lambda$1548,1550 profiles
and found they were too contaminated to fit.
Our fitting results are given in Table~\ref{tab:vpfit_param}.
The COS absorption-line profiles and our VPFIT models are shown 
in Figure~\ref{fig:stack_ebhis}. 

\subsection{\HI\ 21 cm Observations} \label{subsec:hi}

A key parameter in our analysis is the \HI\ column density along 
the two Sextans B CGM sightlines.
We consider several available \HI\ 21 cm datasets.
First we look at single-dish data from the the EBHIS survey 
\citep[10.8\arcmin\ resolution;][]{winkel2016} and
LAB survey \citep[36\arcmin\ resolution;][]{kalberla2005}.
In the top panel of Figure~\ref{fig:stack_ebhis} we include the 
EBHIS or LAB profiles in each direction.
We measured the \HI\ column density in the Sextans B component by
integrating the brightness temperature profile over the 
LSR velocity range 258--348\kms, using the standard relation
$N$(\HI)=1.823$\times10^{18}$\sqcm $\int_{\rm{v_{min}}}^{\rm{v_{max}}}T\rm{_B d}v$,
appropriate for optically-thin gas.

These single-dish profiles are useful for showing the velocity extent of 
the high-velocity \HI\ emission in these directions, but they are subject to large
beam-smearing errors, given the location of the AGN sightlines just off the edge of 
the \HI\ disk of Sextans B (see~Figure~\ref{fig:map}).
Therefore, the LAB and EBHIS measurements do not provide precise 
indications of the \HI\ column along the pencil-beam AGN sightlines.
For that purpose, we use high-resolution interferometric 21 cm data from 
KAT-7 \citep[see Figure 1; 4.4\arcmin$\times$3.2\arcmin\ beam;][]{namumba2018}
and the VLA \citep[LITTLE THINGS survey; 6\arcsec\ beam;][]{hunter2012}, 
which give the \HI\ column-density profiles as a 
function of impact parameter around Sextans B. 
At 4\,kpc, the impact parameter of \jshort, the KAT-7 and VLA profiles
both give the same \HI\ column, $N$(\HI)=$10^{18.0\pm0.2}$\sqcm,
so we adopt this value in our analysis.
At 8\,kpc, the impact parameter of \jtshort, the
\HI\ column density is not well constrained by either the KAT-7 or VLA data; 
we adopt a broad range $N$(\HI)$\sim10^{16-17}$\sqcm, 
based on a linear extrapolation of the KAT-7 profile from 0--4 kpc out to 8 kpc,
using the contours shown in Figure~5 of \citet{namumba2018}.
All the \HI\ measurements are summarized in Table~\ref{tab:vpfit_param}.

\section{Results and Discussion} \label{sec:results}

\begin{figure*}[!ht]
 \includegraphics[width=1.0\textwidth]{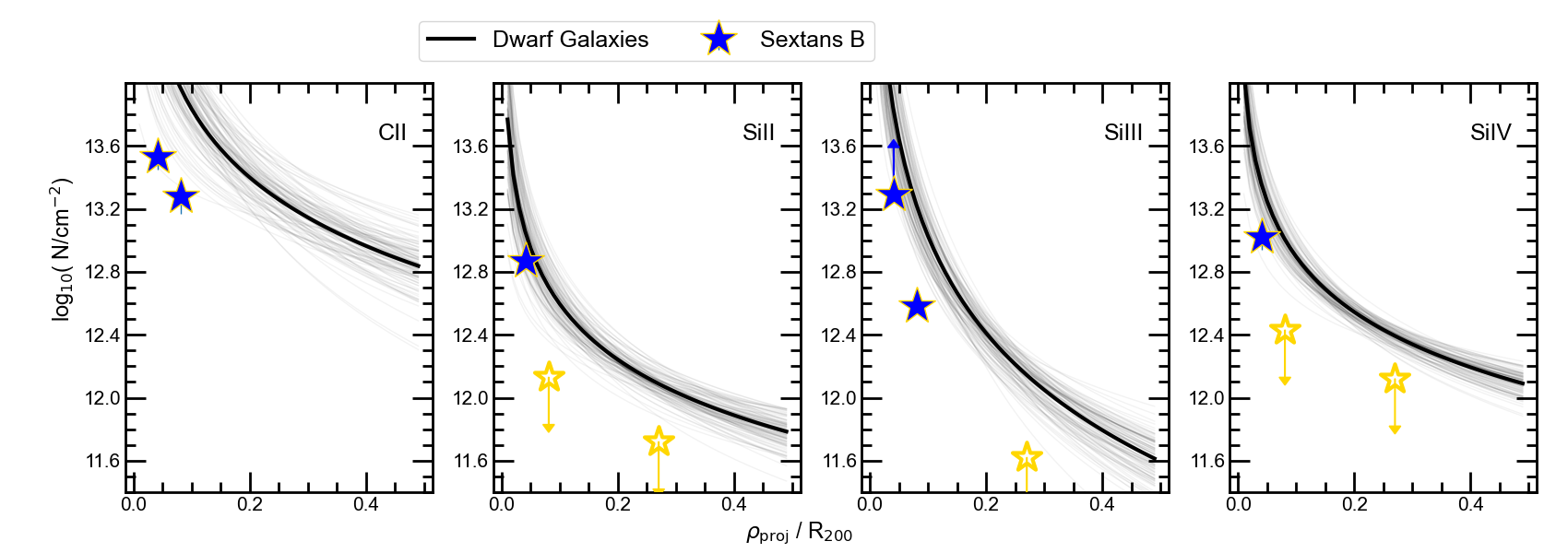}
 \caption{Radial profiles (column density vs. normalized impact parameter)
 of four ions in the Sextans B CGM compared to the average CGM profiles 
from \citetalias{zheng2024} from nearby dwarf galaxies (gray).
Sextans B detections are shown with blue stars, and non-detections 
(upper limits on log\,$N$) are shown with yellow stars. Column densities 
from saturated lines are shown as lower limits. 
The data points at $\rho_{\rm proj}/R_{200}$=0.28 (PG1001+054 sightline) 
are taken from \citetalias{zheng2024}.}
 \label{fig:radial} 
\end{figure*}

\subsection{Detection of Sextans B CGM} \label{subsec:detection}

Toward \jshort\ (at $\rho=4.1$\,kpc), an absorption component at the Sextans B velocity 
($v_{\rm LSR}$=295\kms) is seen in \CII\ $\lambda$1334, 
\SiII\ $\lambda\lambda$1260,1193,1190, \SiIII\ $\lambda$1206, 
and \SiIV\ $\lambda\lambda$1392,1402. 
Two weaker components at 275 and 325\kms\ are suggested by the \SiIII\ $\lambda$1206
\SiII\ $\lambda$1260 and $\lambda$1193 profiles and are included in our VPFIT models.
In our analysis, we adopt the total column density summed across the three 
\SiII\ and \SiIII\ components (see Table~\ref{tab:vpfit_param}), as this allows a 
direct comparison with the \HI,
which is integrated across the entire Sextans B velocity range.

Toward \jtshort\ (at $\rho=8.0$\,kpc), absorption at $v_{\rm LSR}$=295\kms\ is seen in 
\CII\ $\lambda$1334 and \SiIII\ $\lambda$1206, with \SiII\ and \SiIV\ not detected at 
significant levels. The \CIV\ $\lambda$1548 and $\lambda$1550 profiles are blended at 
$v_{\rm LSR}<300$\kms, so we only analyze the absorption at $v_{\rm LSR}>300$\kms; 
the \CIV\ profiles are consistent with the \CII\ and \SiIII\ in this range.
The 295\kms\ component in \CII\ and \SiIII\ matches the systemic velocity of Sextans B 
and the component seen in the \jshort\ sightline, so we identify this as the Sextans B
CGM component. 

Our Sextans B identification for the 295\kms\ component toward \jtshort\ replaces 
the earlier classification of this component
as arising from an extension to the Leading Arm of the Magellanic Stream 
\citep{fox2014, fox2018}. The Leading Arm lies at high positive velocities
of 200--300\kms\ \citep{bq2013} in a nearby region of the sky,
at slightly higher longitudes of 260--310\degree\ than Sextans B at 233\degree.
This earlier work did not consider the halos of Local Group galaxies as 
potential explanations for absorption features, and therefore did not 
consider a Sextans B origin. Our reclassification follows the recent realization 
that the CGMs of LG galaxies can produce absorbers that mimic foreground gas in the 
Magellanic System \citep[][]{zheng2019, kim2024}.


\subsection{Apparent Column Density Profiles} \label{subsec:nav}

To further explore the relationship between the absorption seen in different ions
in the Sextans B, we calculate apparent column density (ACD) profiles using the 
apparent optical depth method \citep{ss91} The ACD profiles show the column 
density in each pixel through the line profile. The ACD is calculated as
$N_a(v)=3.768\times10^{14}(f\lambda)^{-1}\tau(v)$
ions cm$^{-2}$\,(km\,s$^{-1}$)$^{-1}$,
where the apparent optical depth (AOD) is $\tau(v)$=ln$[F_c(v)/F(v)]$
and $f$ is the oscillator strength of the transition, taken from \citet{morton2003} 
and \citet{cashman2017}, 
and $F(v)$ and $F_c(v)$ are the flux and continuum level, respectively,
as a function of velocity.
The errors on the ACD are calculated by propagating the errors on the flux following
\citet{sembach1992}; the COS flux errors include photon noise, flat-field uncertainties, 
and background uncertainties \citep{cosdhb2022}.

The ACD profiles allow for a linear comparison of the absorption in 
different ions, which provides key information on whether those ions are co-spatial 
in the Sextans B CGM. We normalize the ACD profile of each ion to its peak value,
to allow the profile of different ions to be compared.
The ACD profiles are given in Figure~\ref{fig:acd}. They reveal that toward \jshort,
\SiII, \SiIII, and \SiIV\ generally show consistent (self-similar) profiles, although 
there are slight differences between \CII\ and \SiII\ in the velocity 
range 300--340\kms, and between \SiII\ and \SiIII\ at 270--290\kms.
These differences could be caused by sub-structure or unidentified blends.
Notwithstanding these differences, the general similarity of the low ions 
profiles suggests they are likely to be co-spatial,
and can be modeled with a single-phase {\it Cloudy} simulation.
Toward \jtshort, \CII\ and \SiIII\ match each other well at 290\kms,
as does the unblended \CIV\ absorption at $v_{\rm LSR}>300$\kms.
The \CIV\ absorption profiles at $v_{\rm LSR}<300$\kms\ are blended 
and not included in the analysis.
In summary, the ACD analysis finds that in both sightlines,
the ions detected (\SiII, \SiIII, and \SiIV, and \CII) 
have kinematics that are broadly consistent with existing in a single phase.

\subsection{Radial Profile of Sextans B CGM} \label{subsec:radial}

The radial profile of the Sextans B CGM (column density vs. impact parameter)
is shown in Figure~\ref{fig:radial} for \CII, \SiII, \SiIII, and \SiIV.
The profiles include the two inner sightlines from our new analysis 
(\jshort\ and \jtshort) and the 
third sightline (PG1011-040) from \citetalias{zheng2024}. 
To compare the Sextans B profiles with the average dwarf CGM profile, 
we used the sample of nearby dwarf galaxies from \citetalias{zheng2024},
but removed the three Sextans B sightlines before forming the average.
To derive the average CGM profile, we applied a censored regression algorithm 
implemented in PyMC3 to fit a power-law model, following the procedure described 
in Appendix C of \citetalias{zheng2024}. This comparison sample (which contains 44 dwarf galaxies)
has a median halo mass $10^{10.75}$\msun, compared to $10^{10.52}$\msun\ for Sextans B.

The plot shows that all four ions in our Sextans B dataset 
(\CII, \SiII, \SiIII, and \SiIV) fall below the comparison sample,
by $\approx$0.3 dex for the silicon lines and $\approx$0.6 dex for \CII, 
but otherwise show a standard declining radial profile. 
The finding that the Sextans B data falls below the average dwarf CGM
is consistent with the low halo mass and low metallicity of Sextans B,
both of which would serve to lower the metal column densities at a given 
impact parameter and therefore lower the radial profile.

\subsection{Cloudy Photoionization Models} \label{subsec:cloudy}

\begin{figure*}[!ht]
\includegraphics[width=1.0\textwidth]{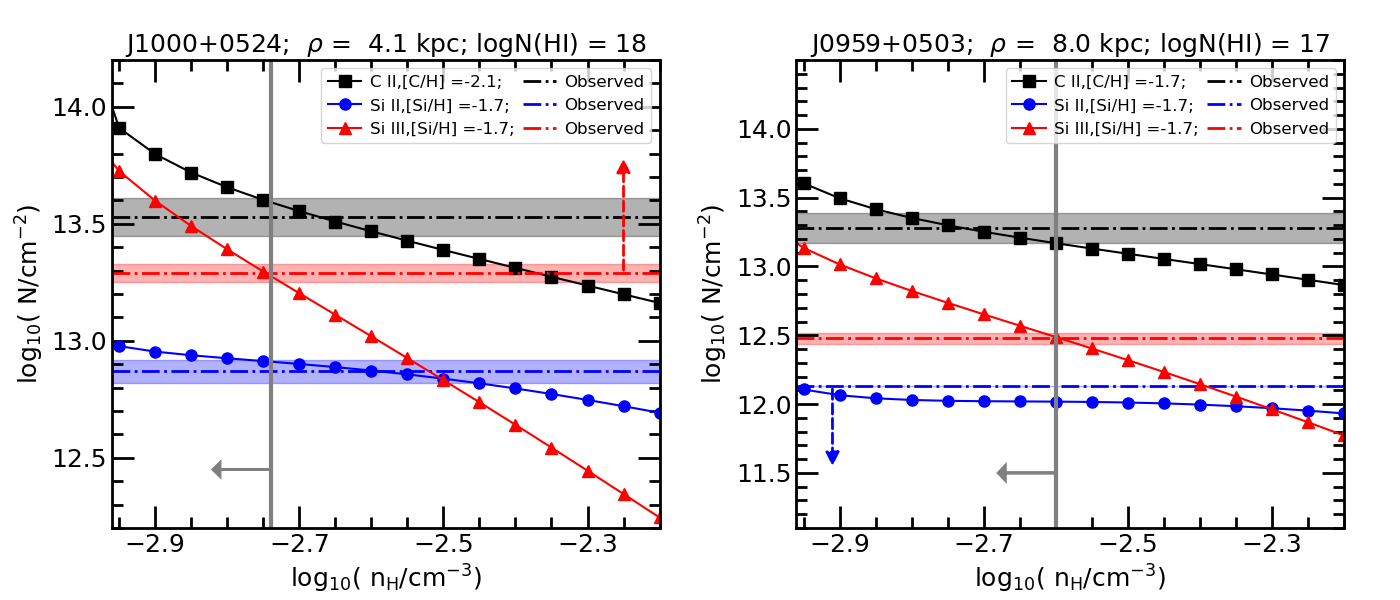}
\caption{Results of \emph{Cloudy} photoionization analysis showing the column
density of \CII, \SiII, and \SiIII\ against gas density $n_{\rm H}$, 
assuming log\,$N$(\HI)=18.0 and 17.0 for 
the \jshort\ (left column) and \jtshort\ (right column) sightlines, respectively.
The model predictions are shown with connected symbols, and the observations 
are shown as colored horizontal bars. Saturated lines give lower limits on 
log\,$N$, and non-detections give upper limits.
The best-fit values of log\,$n_{\rm H}<-2.74$ and $<-2.60$ derived from
the observed \SiIII/\SiII\ ratios 
are shown as the vertical gray lines. 
These models yield the gas-phase silicon and carbon
abundances in the Sextans B CGM in each sightline, as annotated in the legend.}
\label{fig:cloudy} 
\end{figure*}

Using photoionization modeling, we can constrain the physical conditions
and chemical abundances 
in the Sextans B CGM. To do this, we modeled the measured column densities 
of the low ions and the \HI\ in Sextans B using the {\it Cloudy} radiative 
transfer code \citep[v25;][]{gunasekera2025}. 
We assumed the UV background at $z$=0 from \citet{khaire2019} as the 
incident radiation field, which is appropriate since Sextans B lies 
far ($>$1 Mpc) from ionizing radiation from the 
Milky Way \citep{fox2005} and M31 \citep{lehner2025}, 
the dominant sources of ionizing photons in the Local Group.
For the inner sightline (\jshort) we used the \HI\ column density 
measured from the KAT-7 and VLA data 
(log\,$N$(\HI)=18.0; see Section~\ref{subsec:hi})
as the stopping criterion.
For the outer sightline (\jtshort) we ran two sets of models, with
log\,$N$(\HI)=16.0 and 17.0, as $N$(\HI) is not well constrained observationally.

For each sightline 
we ran a grid of models with gas density log\,($n_{\rm H}$/cm$^{-3}$) 
varying from $-3.0$ to $-1.0$ in 0.05\,dex intervals, and metallicity [Z/H] 
varying from $-$3.0 to 0.0 in 0.1\,dex intervals.
The density is related to the ionization parameter $U$ 
and the ionizing photon density $n_\gamma$ by $U=n_\gamma/n_{\rm H}$. 
For each sightline, we first determine the best-fit value of $n_{\rm H}$ by
matching the observed \SiIII/\SiII\ ratio, which is independent of metallicity.
Our observations provide a \emph{lower limit} on the \SiIII/\SiII\ ratio 
in the Sextans B CGM
in both sightlines, because of saturation in \SiIII\ 1206 toward \jshort,
and a non-detection of \SiII\ 1260 toward \jtshort; 
this leads to an \emph{upper limit} on the allowed values of $n_{\rm H}$.
We then determine the best-fit gas-phase silicon and carbon abundances,
[Si/H] and [C/H], by matching the absolute 
column densities of \SiII\ and \SiIII\ (for Si) and \CII\ (for C) 
at the best-fit $n_{\rm H}$.

The results of the {\it Cloudy} modeling are shown in Figure~\ref{fig:cloudy}.
For the \jshort\ sightline at 4 kpc (where the \HI\ column is well constrained) we derive 
log\,$n_{\rm H}<-2.74$, [Si/H]=$-1.7\pm0.2$, and [C/H]=$-2.1\pm0.2$.
For the \jtshort\ sightline at 8 kpc we derive 
log\,$n_{\rm H}<-2.60$, [Si/H]=$-1.7\pm0.2$, and [C/H]=$-1.7\pm0.2$ from the model
with log\,$N$(\HI)=17.
The finding that the same value of [Si/H] is found for the 
two sightlines is encouraging, and supports our choice of log\,$N$(\HI)=17 
as the appropriate \HI\ column for the \jtshort\ model. 
However, we cannot rule out solutions in the \jtshort\ sightline
with lower $N$(\HI) and higher [Si/H].
The model with log\,$N$(\HI)=16 gives [Si/H]=$-$0.7; although this model can 
explain the UV column densities, we consider it unrealistic as metallicities
this high are not seen anywhere in the Sextans B system, either in stars or gas.

The silicon abundance provides an estimate of the overall metallicity 
in the Sextans B CGM if there is no dust depletion. Our modeling therefore 
provides the first constraint on the metallicity of the Sextans B CGM.
Our value [Si/H]=$-1.7\pm0.2$ (2\% solar) from the \jshort\ sightline
is similar to the stellar metallicity of 
[Fe/H]=$-$1.9 to $-$1.6 in Sextans B \citep{vdb2000, bellazzini2014} and
is on the low end of the distribution of CGM metallicities
at low redshift, which peaks near [Z/H]=$-$1.0 \citep{prochaska2017}.
The low metallicity is not surprising given the low mass of Sextans B and the 
mass-metallicity relation.
To our knowledge, the only other gas-phase metallicities in the Local Group 
as low as this are the three compact high-velocity clouds measured by \citet{ashley2024} 
with oxygen abundances of 2--4\% solar.
Our measurement is only the third metallicity measurement
in the cool low-ion CGM of a dwarf galaxy at any redshift \citep{zahedy2021, johnson2026}.

The finding that [C/H] is 0.4\,dex lower than [Si/H] in the Sextans B CGM 
explains why the \CII\ data points in Figure~\ref{fig:radial} lie further below the 
average CGM profile than the \SiII\ and \SiIII\ data points.
This under-abundance of carbon is unlikely to be a dust depletion effect, 
since carbon depletes less than silicon in the ISM
\citep[e.g.][]{1996ARA&A..34..279S, jenkins2009}, 
so if dust were the cause of the difference [C/H] would
be \emph{higher} than [Si/H]. Instead this may be a nucleosynthesis effect, 
in which silicon (an $\alpha$-element produced in massive stars) has been 
preferentially produced over carbon (dominated by the CNO cycle in less massive 
stars) over the star formation history of Sextans B, resulting in a non-solar
ratio between the two elements at the current time.

\subsection{CGM Silicon Mass and Gas Mass} \label{subsec:metals}

Since our Sextans B CGM observations cover three successive ionization states 
of silicon (\SiII, \SiIII, \SiIV), we can calculate the total column 
density of silicon in each sightline by summing over the observed 
ionization states, $N_{\rm Si}=N$(\SiII)+$N$(\SiIII)+$N$(\SiIV).
This approach has been followed in several CGM studies of LG galaxies 
\citep[e.g.][]{lehner2015, richter2016, fox2019, zheng2019, zheng2020} because it 
allows the total silicon mass
(and CGM mass) to be calculated even in the absence of \emph{Cloudy} models.
At the expected temperature for the CGM of dwarf galaxies, the contribution from 
higher ionization states is expected to be small \citep{gnat2007}.
We do not include upper limits for non-detected lines, so for the \jtshort\ 
sightline $N_{\rm Si}$=$N$(\SiIII).

To convert the total Si columns to a CGM Si mass, we integrate over 
the radial profile of the CGM, 
$M_{\rm Si}^{\rm CGM}(r)=m_{\rm Si}\int N_{\rm Si}(r) f_{\rm cov}2\pi r dr $, 
where $f_{\rm cov}$ is the sky covering fraction, $m_{\rm Si}$ is the mass of a silicon atom, and we assume a symmetric
(circular) on-sky CGM projection. To evaluate this expression, we assume
$f_{\rm cov}$=1 ;
in reality the CGM covering fraction declines with radius, 
but given the very small impact parameters (4 and 8 kpc) of the two sightlines
$f_{\rm cov}$=1 is a reasonable assumption.
Since we have two sightlines, we split the integral into two concentric 
annular regions at 
$\rho\!<\!4$\,kpc and  $4\!<\!\rho\!<\!8$\,kpc. 
Assuming a constant column density within each region,
the Si mass can be written as a sum over two parts \citep{zheng2020}:

\begin{equation}
M_{\rm Si}^{\rm CGM}(r)=m_{\rm Si}\pi\sum_{i=1}^{i=2}  (r_i^2 - r_{i-1}^2)N_{\rm Si}(r_i).
\end{equation} 

This technique could be extended out to the PG 1001+054 sightline at 27 kpc,
but since there are no CGM detections in this sightline (just upper limits), 
we adopt the more conservative
approach of only reporting the CGM mass in the regions with detections.
The results are given in Table~\ref{tab:mass}.
The total Si mass out to 8 kpc is $5.5\times10^2$\msun.
Given the stellar mass of Sextans B \citepalias[$4.0\times10^7$\msun;][]{zheng2024}
and the silicon yield from stellar nucleosynthesis \citep[0.003;][]{zheng2020}, 
the total Si mass produced in stars in Sextans B is $\approx1.2\times10^5$\msun.
Our CGM observations within 8 kpc account for $\approx$0.5\% of this total. 

Given the CGM Si mass, we can calculate the total cool CGM mass by 
correcting for the Si abundance,
$M_{\rm cool}^{\rm CGM}=\mu M_{\rm Si}^{\rm CGM}/[(m_{\rm Si}/m_{\rm H})({\rm Si/H})^{\rm CGM}]$,
where $\mu$ is the mean molecular weight.
We adopt a CGM metallicity [Si/H]$^{\rm CGM}$=$-$1.7 
based on the 
\emph{Cloudy} results toward the inner sightline \jshort\
(see Section~\ref{subsec:cloudy}), and use $\mu=1.4$, giving a 
cool CGM mass within 8 kpc of 
$\approx3.9\times10^{7}$\msun, comparable to 
the ISM \HI\ mass of $4.1\times10^7$\msun\ \citep{hunter2012}
and the stellar mass of $4.0\times10^7$\msun\ \citepalias{zheng2024} in Sextans B.
The CGM mass is a lower limit since integrating to higher radii 
would increase the mass estimate.
While further CGM sightlines are needed for more precise estimates,
these numbers 
suggest that the Sextans B CGM is the galaxy's 
dominant baryon reservoir, as is true for other dwarf-galaxy CGMs 
\citep[Z24;][]{mishran2024, piacitelli2025} and the CGM in general \citep{werk2014}.

\section{Summary} \label{sec:summary}

We have conducted a UV absorption-line analysis of the CGM of 
the dwarf irregular galaxy Sextans B
using HST/COS observations of the AGN \jshort\ and \jtshort.
Sextans B lies in an isolated environment on the edge of the LG
where galaxy interactions are expected to be minimal.
The two sightlines probe the Sextans B CGM at small impact parameters 
of 4.1 and 8.0 kpc ($\approx$0.04 and 0.08\rvir), respectively.

We detect the Sextans B CGM in low-ion (\SiII, \SiIII, \CII) and high-ion 
(\SiIV) absorption. All four ions show a declining radial profile that falls  
$\approx$0.3--0.6\,dex below the average CGM profile of nearby dwarf galaxies, 
consistent with the low halo mass and low metallicity of Sextans B.

Using {\it Cloudy} photoionization modeling of the column densities measured
toward \jshort\ (where the \HI\ column density is well constrained),
we derive a gas density log\,$n_{\rm H}<-2.74$ and a low silicon and carbon 
abundance [Si/H]=$-1.7\pm0.2$ (2\% solar) and [C/H]=$-2.1\pm0.2$  
in the Sextans B CGM. In the absence of dust, the silicon abundance provides a 
measurement of the CGM metallicity; this value matches the stellar metallicity 
of Sextans B and is among the lowest gas-phase metallicities measured in the Local Group. 
The low-ion CGM within 8 kpc has a gas mass of 
$\approx4\times10^{7}$\msun, comparable to the stellar mass 
and \HI\ mass in Sextans B.

Our study places constraints on the physical and chemical properties of a 
gaseous halo in a quiescent low-density environment where ram-pressure effects 
are expected to be minimal.  
Our results expand the number of LG galaxies that have multiple UV CGM detections and 
therefore observational constraints on the CGM radial profile. Sextans B joins the 
LMC, M31, IC1613, and Sextans A in this category.
These profiles provide key observational constraints on
the diffuse gas content of the Local Group.

\vspace{0.6cm}
{\it Acknowledgments:}
We thank Brenda Namumba for giving us permission to reproduce the KAT-7 data in Figure 1,
Louise Breuval for valuable discussions on the Sextans B metallicity, and Sean Johnson for helpful comments. We thank the referee for a constructive report.
Support for program 17209 was provided by NASA through a grant from 
the Space Telescope Science Institute, which is operated by the 
Association of Universities for Research in Astronomy, Inc., under NASA contract NAS5-26555. 


The HST data presented in this article were obtained from the Mikulski Archive for Space Telescopes (MAST) at the Space Telescope Science Institute. The specific observations analyzed can be accessed via \dataset[DOI: 10.17909/9q7e-js22]{https://doi.org/10.17909/9q7e-js22}.

\facilities{HST/COS \citep{green2012}, EBHIS \citep{winkel2016}, LAB \citep{kalberla2005}, KAT-7 \citep{foley2016}, VLA \citep{hunter2012}.}
\software{VPFIT \citep{Carswell2014}, HASP \citep{debes2024}, \emph{Cloudy} \citep{gunasekera2025}.}

\bibliography{Reference}{}

@ARTICLE{green2012,
       author = {{Green}, James C. and {Froning}, Cynthia S. and {Osterman}, Steve and {Ebbets}, Dennis and {Heap}, Sara H. and {Leitherer}, Claus and {Linsky}, Jeffrey L. and {Savage}, Blair D. and {Sembach}, Kenneth and {Shull}, J. Michael and {Siegmund}, Oswald H.~W. and {Snow}, Theodore P. and {Spencer}, John and {Stern}, S. Alan and {Stocke}, John and {Welsh}, Barry and {B{\'e}land}, St{\'e}phane and {Burgh}, Eric B. and {Danforth}, Charles and {France}, Kevin and {Keeney}, Brian and {McPhate}, Jason and {Penton}, Steven V. and {Andrews}, John and {Brownsberger}, Kenneth and {Morse}, Jon and {Wilkinson}, Erik},
        title = "{The Cosmic Origins Spectrograph}",
      journal = {\apj},
         year = 2012,
        month = jan,
       volume = {744},
       number = {1},
          eid = {60},
        pages = {60},
          doi = {10.1088/0004-637X/744/1/6010.1086/141956},
archivePrefix = {arXiv},
       eprint = {1110.0462},
 primaryClass = {astro-ph.IM},
       adsurl = {https://ui.adsabs.harvard.edu/abs/2012ApJ...744...60G}
}

@ARTICLE{mishra2024,
       author = {{Mishra}, Sapna and {Fox}, Andrew J. and {Krishnarao}, Dhanesh and {Lucchini}, Scott and {D'Onghia}, Elena and {Cashman}, Frances H. and {Barger}, Kathleen A. and {Lehner}, Nicolas and {Tumlinson}, Jason},
        title = "{The Truncated Circumgalactic Medium of the Large Magellanic Cloud}",
      journal = {\apjl},
         year = 2024,
        month = dec,
       volume = {976},
       number = {2},
          eid = {L28},
        pages = {L28},
          doi = {10.3847/2041-8213/ad8b9d},
archivePrefix = {arXiv},
       eprint = {2410.11960},
 primaryClass = {astro-ph.GA},
       adsurl = {https://ui.adsabs.harvard.edu/abs/2024ApJ...976L..28M}
}

@ARTICLE{fox2014,
       author = {{Fox}, Andrew J. and {Wakker}, Bart P. and {Barger}, Kathleen A. and {Hernandez}, Audra K. and {Richter}, Philipp and {Lehner}, Nicolas and {Bland-Hawthorn}, Joss and {Charlton}, Jane C. and {Westmeier}, Tobias and {Thom}, Christopher and {Tumlinson}, Jason and {Misawa}, Toru and {Howk}, J. Christopher and {Haffner}, L. Matthew and {Ely}, Justin and {Rodriguez-Hidalgo}, Paola and {Kumari}, Nimisha},
        title = "{The COS/UVES Absorption Survey of the Magellanic Stream. III. Ionization, Total Mass, and Inflow Rate onto the Milky Way}",
      journal = {\apj},
         year = 2014,
        month = jun,
       volume = {787},
       number = {2},
          eid = {147},
        pages = {147},
          doi = {10.1088/0004-637X/787/2/147},
archivePrefix = {arXiv},
       eprint = {1404.5514},
 primaryClass = {astro-ph.GA},
       adsurl = {https://ui.adsabs.harvard.edu/abs/2014ApJ...787..147F}
}

@ARTICLE{fox2018,
       author = {{Fox}, Andrew J. and {Barger}, Kathleen A. and {Wakker}, Bart P. and {Richter}, Philipp and {Antwi-Danso}, Jacqueline and {Casetti-Dinescu}, Dana I. and {Howk}, J. Christopher and {Lehner}, Nicolas and {D'Onghia}, Elena and {Crowther}, Paul A. and {Lockman}, Felix J.},
        title = "{Chemical Abundances in the Leading Arm of the Magellanic Stream}",
      journal = {\apj},
         year = 2018,
        month = feb,
       volume = {854},
       number = {2},
          eid = {142},
        pages = {142},
          doi = {10.3847/1538-4357/aaa9bb},
archivePrefix = {arXiv},
       eprint = {1801.06446},
 primaryClass = {astro-ph.GA},
       adsurl = {https://ui.adsabs.harvard.edu/abs/2018ApJ...854..142F}
}

@ARTICLE{zheng2024,
       author = {{Zheng}, Yong and {Faerman}, Yakov and {Oppenheimer}, Benjamin D. and {Putman}, Mary E. and {McQuinn}, Kristen B.~W. and {Kirby}, Evan N. and {Burchett}, Joseph N. and {Telford}, O. Grace and {Werk}, Jessica K. and {Kim}, Doyeon A.},
        title = "{A Comprehensive Investigation of Metals in the Circumgalactic Medium of Nearby Dwarf Galaxies}",
      journal = {\apj},
         year = 2024,
        month = jan,
       volume = {960},
       number = {1},
          eid = {55},
        pages = {55},
          doi = {10.3847/1538-4357/acfe6b},
archivePrefix = {arXiv},
       eprint = {2301.12233},
 primaryClass = {astro-ph.GA},
       adsurl = {https://ui.adsabs.harvard.edu/abs/2024ApJ...960...55Z}
}

@ARTICLE{qu2022,
       author = {{Qu}, Zhijie and {Bregman}, Joel N.},
        title = "{Absorption Line Search through Three Local Group Dwarf Galaxy Halos}",
      journal = {\apj},
         year = 2022,
        month = mar,
       volume = {927},
       number = {2},
          eid = {228},
        pages = {228},
          doi = {10.3847/1538-4357/ac51df},
archivePrefix = {arXiv},
       eprint = {2203.08246},
 primaryClass = {astro-ph.GA},
       adsurl = {https://ui.adsabs.harvard.edu/abs/2022ApJ...927..228Q}
}

@ARTICLE{zheng2020,
       author = {{Zheng}, Yong and {Emerick}, Andrew and {Putman}, Mary E. and {Werk}, Jessica K. and {Kirby}, Evan N. and {Peek}, Joshua},
        title = "{Characterizing the Circumgalactic Medium of the Lowest-mass Galaxies: A Case Study of IC 1613}",
      journal = {\apj},
         year = 2020,
        month = dec,
       volume = {905},
       number = {2},
          eid = {133},
        pages = {133},
          doi = {10.3847/1538-4357/abc875},
archivePrefix = {arXiv},
       eprint = {2010.15645},
 primaryClass = {astro-ph.GA},
       adsurl = {https://ui.adsabs.harvard.edu/abs/2020ApJ...905..133Z}
}

@ARTICLE{zheng2019,
       author = {{Zheng}, Yong and {Putman}, Mary E. and {Emerick}, Andrew and {McQuinn}, Kristen B.~W. and {Werk}, Jessica K. and {Lockman}, Felix J. and {Oppenheimer}, Benjamin D. and {Fox}, Andrew J. and {Kirby}, Evan N. and {Burchett}, Joseph N.},
        title = "{Tentative detection of the circumgalactic medium of the isolated low-mass dwarf galaxy WLM}",
      journal = {\mnras},
         year = 2019,
        month = nov,
       volume = {490},
       number = {1},
        pages = {467-477},
          doi = {10.1093/mnras/stz2563},
archivePrefix = {arXiv},
       eprint = {1909.05407},
 primaryClass = {astro-ph.GA},
       adsurl = {https://ui.adsabs.harvard.edu/abs/2019MNRAS.490..467Z}
}

@ARTICLE{winkel2016,
       author = {{Winkel}, B. and {Kerp}, J. and {Fl{\"o}er}, L. and {Kalberla}, P.~M.~W. and {Ben Bekhti}, N. and {Keller}, R. and {Lenz}, D.},
        title = "{The Effelsberg-Bonn H I Survey: Milky Way gas. First data release}",
      journal = {\aap},
         year = 2016,
        month = jan,
       volume = {585},
          eid = {A41},
        pages = {A41},
          doi = {10.1051/0004-6361/201527007},
archivePrefix = {arXiv},
       eprint = {1512.05348},
 primaryClass = {astro-ph.IM},
       adsurl = {https://ui.adsabs.harvard.edu/abs/2016A&A...585A..41W}
}

@software{Carswell2014,
       author = {{Carswell}, R.~F. and {Webb}, J.~K.},
        title = "{VPFIT: Voigt profile fitting program}",
 howpublished = {Astrophysics Source Code Library, record ascl:1408.015},
         year = 2014,
        month = aug,
          eid = {ascl:1408.015},
       adsurl = {https://ui.adsabs.harvard.edu/abs/2014ascl.soft08015C}
}

@ARTICLE{namumba2018,
       author = {{Namumba}, B. and {Carignan}, C. and {Passmoor}, S.},
        title = "{H I observations of Sextans A and B with the SKA pathfinder KAT-7}",
      journal = {\mnras},
         year = 2018,
        month = jul,
       volume = {478},
       number = {1},
        pages = {487-500},
          doi = {10.1093/mnras/sty1056},
archivePrefix = {arXiv},
       eprint = {1804.07730},
 primaryClass = {astro-ph.GA},
       adsurl = {https://ui.adsabs.harvard.edu/abs/2018MNRAS.478..487N}
}

@BOOK{vdb2000,
       author = {{van den Bergh}, Sidney},
        title = "{The Galaxies of the Local Group}",
         year = 2000,
       adsurl = {https://ui.adsabs.harvard.edu/abs/2000glg..book.....V}
}

@ARTICLE{mcc2012,
       author = {{McConnachie}, Alan W.},
        title = "{The Observed Properties of Dwarf Galaxies in and around the Local Group}",
      journal = {\aj},
         year = 2012,
        month = jul,
       volume = {144},
       number = {1},
          eid = {4},
        pages = {4},
          doi = {10.1088/0004-6256/144/1/4},
archivePrefix = {arXiv},
       eprint = {1204.1562},
 primaryClass = {astro-ph.CO},
       adsurl = {https://ui.adsabs.harvard.edu/abs/2012AJ....144....4M}
}

@ARTICLE{bordoloi2014,
       author = {{Bordoloi}, Rongmon and {Tumlinson}, Jason and {Werk}, Jessica K. and {Oppenheimer}, Benjamin D. and {Peeples}, Molly S. and {Prochaska}, J. Xavier and {Tripp}, Todd M. and {Katz}, Neal and {Dav{\'e}}, Romeel and {Fox}, Andrew J. and {Thom}, Christopher and {Ford}, Amanda Brady and {Weinberg}, David H. and {Burchett}, Joseph N. and {Kollmeier}, Juna A.},
        title = "{The COS-Dwarfs Survey: The Carbon Reservoir around Sub-L* Galaxies}",
      journal = {\apj},
         year = 2014,
        month = dec,
       volume = {796},
       number = {2},
          eid = {136},
        pages = {136},
          doi = {10.1088/0004-637X/796/2/136},
archivePrefix = {arXiv},
       eprint = {1406.0509},
 primaryClass = {astro-ph.GA},
       adsurl = {https://ui.adsabs.harvard.edu/abs/2014ApJ...796..136B}
}

@ARTICLE{prochaska2017,
       author = {{Prochaska}, J. Xavier and {Werk}, Jessica K. and {Worseck}, G{\'a}bor and {Tripp}, Todd M. and {Tumlinson}, Jason and {Burchett}, Joseph N. and {Fox}, Andrew J. and {Fumagalli}, Michele and {Lehner}, Nicolas and {Peeples}, Molly S. and {Tejos}, Nicolas},
        title = "{The COS-Halos Survey: Metallicities in the Low-redshift Circumgalactic Medium}",
      journal = {\apj},
         year = 2017,
        month = mar,
       volume = {837},
       number = {2},
          eid = {169},
        pages = {169},
          doi = {10.3847/1538-4357/aa6007},
archivePrefix = {arXiv},
       eprint = {1702.02618},
 primaryClass = {astro-ph.GA},
       adsurl = {https://ui.adsabs.harvard.edu/abs/2017ApJ...837..169P}
}

@ARTICLE{werk2014,
       author = {{Werk}, Jessica K. and {Prochaska}, J. Xavier and {Tumlinson}, Jason and {Peeples}, Molly S. and {Tripp}, Todd M. and {Fox}, Andrew J. and {Lehner}, Nicolas and {Thom}, Christopher and {O'Meara}, John M. and {Ford}, Amanda Brady and {Bordoloi}, Rongmon and {Katz}, Neal and {Tejos}, Nicolas and {Oppenheimer}, Benjamin D. and {Dav{\'e}}, Romeel and {Weinberg}, David H.},
        title = "{The COS-Halos Survey: Physical Conditions and Baryonic Mass in the Low-redshift Circumgalactic Medium}",
      journal = {\apj},
         year = 2014,
        month = sep,
       volume = {792},
       number = {1},
          eid = {8},
        pages = {8},
          doi = {10.1088/0004-637X/792/1/8},
archivePrefix = {arXiv},
       eprint = {1403.0947},
 primaryClass = {astro-ph.CO},
       adsurl = {https://ui.adsabs.harvard.edu/abs/2014ApJ...792....8W}
}

@ARTICLE{gnat2007,
       author = {{Gnat}, Orly and {Sternberg}, Amiel},
        title = "{Time-dependent Ionization in Radiatively Cooling Gas}",
      journal = {\apjs},
         year = 2007,
        month = feb,
       volume = {168},
       number = {2},
        pages = {213-230},
          doi = {10.1086/509786},
archivePrefix = {arXiv},
       eprint = {astro-ph/0608181},
 primaryClass = {astro-ph},
       adsurl = {https://ui.adsabs.harvard.edu/abs/2007ApJS..168..213G}
}

@ARTICLE{kniazev2005,
       author = {{Kniazev}, Alexei Y. and {Grebel}, Eva K. and {Pustilnik}, Simon A. and {Pramskij}, Alexander G. and {Zucker}, Daniel B.},
        title = "{Spectrophotometry of Sextans A and B: Chemical Abundances of H II Regions and Planetary Nebulae}",
      journal = {\aj},
         year = 2005,
        month = oct,
       volume = {130},
       number = {4},
        pages = {1558-1573},
          doi = {10.1086/432931},
archivePrefix = {arXiv},
       eprint = {astro-ph/0502562},
 primaryClass = {astro-ph},
       adsurl = {https://ui.adsabs.harvard.edu/abs/2005AJ....130.1558K}
}

@MISC{debes2024,
       author = {{Debes}, John and {Sankrit}, Ravi and {Fischer}, Travis and {Frazer}, Elaine and {Hirschauer}, Alec and {Rowlands}, Kate and {Burger}, Matthew and {Swaters}, Robert and {Jedrzejewski}, Robert and {Gomez}, Sierrra and {Dos Santos}, Leonardo and {Hernandez}, Svea and {Miller}, Lauren and {Payne}, Anna and {Rafelski}, Marc and {Wevers}, Thomas and {Anderson}, Sara and {Bair}, Tom and {Bello}, Kathryn and {Carlberg}, Joleen and {Charlow}, Brian and {Cortese}, Andrew and {Dencheva}, Nadia and {Ellis}, Tracy and {Falk}, Ben and {Fleming}, Scott and {Forshay}, Peter and {Gilani}, Syed and {Hall}, Patty and {Kimball}, Tim and {Kelley}, Talya and {Kidwell}, Richard and {Kotler}, Jenn and {Kovacs}, Aiden and {James}, Bethan and {Rahmani}, Christopher and {Rodriguez}, David and {Roman-Duval}, Julia and {Soderblom}, David and {Sherbert}, Lisa and {Welty}, Dan and {Wolfe}, David},
        title = "{The Hubble Advanced Spectral Product (HASP) Program}",
 howpublished = {Instrument Science Report COS 2024-01, 31 pages},
         year = 2024,
        month = jan,
        pages = {1},
       adsurl = {https://ui.adsabs.harvard.edu/abs/2024cos..rept....1D}
}

@ARTICLE{bellazzini2014,
       author = {{Bellazzini}, M. and {Beccari}, G. and {Fraternali}, F. and {Oosterloo}, T.~A. and {Sollima}, A. and {Testa}, V. and {Galleti}, S. and {Perina}, S. and {Faccini}, M. and {Cusano}, F.},
        title = "{The extended structure of the dwarf irregular galaxies Sextans A and Sextans B. Signatures of tidal distortion in the outskirts of the Local Group}",
      journal = {\aap},
         year = 2014,
        month = jun,
       volume = {566},
          eid = {A44},
        pages = {A44},
          doi = {10.1051/0004-6361/201423659},
archivePrefix = {arXiv},
       eprint = {1404.1697},
 primaryClass = {astro-ph.GA},
       adsurl = {https://ui.adsabs.harvard.edu/abs/2014A&A...566A..44B}
}

@ARTICLE{kalberla2005,
       author = {{Kalberla}, P.~M.~W. and {Burton}, W.~B. and {Hartmann}, Dap and {Arnal}, E.~M. and {Bajaja}, E. and {Morras}, R. and {P{\"o}ppel}, W.~G.~L.},
        title = "{The Leiden/Argentine/Bonn (LAB) Survey of Galactic HI. Final data release of the combined LDS and IAR surveys with improved stray-radiation corrections}",
      journal = {\aap},
         year = 2005,
        month = sep,
       volume = {440},
       number = {2},
        pages = {775-782},
          doi = {10.1051/0004-6361:20041864},
archivePrefix = {arXiv},
       eprint = {astro-ph/0504140},
 primaryClass = {astro-ph},
       adsurl = {https://ui.adsabs.harvard.edu/abs/2005A&A...440..775K}
}

@ARTICLE{tully2006,
       author = {{Tully}, R. Brent and {Rizzi}, L. and {Dolphin}, A.~E. and {Karachentsev}, I.~D. and {Karachentseva}, V.~E. and {Makarov}, D.~I. and {Makarova}, L. and {Sakai}, S. and {Shaya}, E.~J.},
        title = "{Associations of Dwarf Galaxies}",
      journal = {\aj},
         year = 2006,
        month = aug,
       volume = {132},
       number = {2},
        pages = {729-748},
          doi = {10.1086/505466},
archivePrefix = {arXiv},
       eprint = {astro-ph/0603380},
 primaryClass = {astro-ph},
       adsurl = {https://ui.adsabs.harvard.edu/abs/2006AJ....132..729T}
}

@ARTICLE{vdb1999,
       author = {{van den Bergh}, Sidney},
        title = "{The Nearest Group of Galaxies}",
      journal = {\apjl},
         year = 1999,
        month = jun,
       volume = {517},
       number = {2},
        pages = {L97-L99},
          doi = {10.1086/312044},
archivePrefix = {arXiv},
       eprint = {astro-ph/9904425},
 primaryClass = {astro-ph},
       adsurl = {https://ui.adsabs.harvard.edu/abs/1999ApJ...517L..97V}
}

@ARTICLE{bellazzini2013,
       author = {{Bellazzini}, M. and {Oosterloo}, T. and {Fraternali}, F. and {Beccari}, G.},
        title = "{Dwarfs walking in a row. The filamentary nature of the NGC 3109 association}",
      journal = {\aap},
         year = 2013,
        month = nov,
       volume = {559},
          eid = {L11},
        pages = {L11},
          doi = {10.1051/0004-6361/201322744},
archivePrefix = {arXiv},
       eprint = {1310.6365},
 primaryClass = {astro-ph.CO},
       adsurl = {https://ui.adsabs.harvard.edu/abs/2013A&A...559L..11B}
}

@ARTICLE{weisz2011,
       author = {{Weisz}, Daniel R. and {Dalcanton}, Julianne J. and {Williams}, Benjamin F. and {Gilbert}, Karoline M. and {Skillman}, Evan D. and {Seth}, Anil C. and {Dolphin}, Andrew E. and {McQuinn}, Kristen B.~W. and {Gogarten}, Stephanie M. and {Holtzman}, Jon and {Rosema}, Keith and {Cole}, Andrew and {Karachentsev}, Igor D. and {Zaritsky}, Dennis},
        title = "{The ACS Nearby Galaxy Survey Treasury. VIII. The Global Star Formation Histories of 60 Dwarf Galaxies in the Local Volume}",
      journal = {\apj},
         year = 2011,
        month = sep,
       volume = {739},
       number = {1},
          eid = {5},
        pages = {5},
          doi = {10.1088/0004-637X/739/1/5},
archivePrefix = {arXiv},
       eprint = {1101.1093},
 primaryClass = {astro-ph.CO},
       adsurl = {https://ui.adsabs.harvard.edu/abs/2011ApJ...739....5W}
}

@ARTICLE{fox2019,
       author = {{Fox}, Andrew J. and {Richter}, Philipp and {Ashley}, Trisha and {Heckman}, Timothy M. and {Lehner}, Nicolas and {Werk}, Jessica K. and {Bordoloi}, Rongmon and {Peeples}, Molly S.},
        title = "{The Mass Inflow and Outflow Rates of the Milky Way}",
      journal = {\apj},
         year = 2019,
        month = oct,
       volume = {884},
       number = {1},
          eid = {53},
        pages = {53},
          doi = {10.3847/1538-4357/ab40ad},
archivePrefix = {arXiv},
       eprint = {1909.05561},
 primaryClass = {astro-ph.GA},
       adsurl = {https://ui.adsabs.harvard.edu/abs/2019ApJ...884...53F}
}

@ARTICLE{munshi2021,
       author = {{Munshi}, Ferah and {Brooks}, Alyson M. and {Applebaum}, Elaad and {Christensen}, Charlotte R. and {Quinn}, T. and {Sligh}, Serena},
        title = "{Quantifying Scatter in Galaxy Formation at the Lowest Masses}",
      journal = {\apj},
         year = 2021,
        month = dec,
       volume = {923},
       number = {1},
          eid = {35},
        pages = {35},
          doi = {10.3847/1538-4357/ac0db6},
archivePrefix = {arXiv},
       eprint = {2101.05822},
 primaryClass = {astro-ph.GA},
       adsurl = {https://ui.adsabs.harvard.edu/abs/2021ApJ...923...35M}
}

@ARTICLE{gunasekera2025,
       author = {{Gunasekera}, Chamani M. and {van Hoof}, Peter A.~M. and {Dehghanian}, Maryam and {Chakraborty}, Priyanka and {Shaw}, Gargi and {Bianchi}, Stefano and {Chatzikos}, Marios and {Tsujimoto}, Masahiro and {Ferland}, Gary J.},
        title = "{The 2025 Release of Cloudy}",
      journal = {arXiv e-prints},
         year = 2025,
        month = aug,
          eid = {arXiv:2508.01102},
        pages = {arXiv:2508.01102},
          doi = {10.48550/arXiv.2508.01102},
archivePrefix = {arXiv},
       eprint = {2508.01102},
 primaryClass = {astro-ph.GA},
       adsurl = {https://ui.adsabs.harvard.edu/abs/2025arXiv250801102G}
}

@ARTICLE{shaya2013,
       author = {{Shaya}, Ed J. and {Tully}, R. Brent},
        title = "{The formation of Local Group planes of galaxies}",
      journal = {\mnras},
         year = 2013,
        month = dec,
       volume = {436},
       number = {3},
        pages = {2096-2119},
          doi = {10.1093/mnras/stt1714},
archivePrefix = {arXiv},
       eprint = {1307.4297},
 primaryClass = {astro-ph.CO},
       adsurl = {https://ui.adsabs.harvard.edu/abs/2013MNRAS.436.2096S}
}

@ARTICLE{khaire2019,
       author = {{Khaire}, Vikram and {Srianand}, Raghunathan},
        title = "{New synthesis models of consistent extragalactic background light over cosmic time}",
      journal = {\mnras},
         year = 2019,
        month = apr,
       volume = {484},
       number = {3},
        pages = {4174-4199},
          doi = {10.1093/mnras/stz174},
archivePrefix = {arXiv},
       eprint = {1801.09693},
 primaryClass = {astro-ph.GA},
       adsurl = {https://ui.adsabs.harvard.edu/abs/2019MNRAS.484.4174K}
}

@ARTICLE{ss91,
       author = {{Savage}, Blair D. and {Sembach}, Kenneth R.},
        title = "{The Analysis of Apparent Optical Depth Profiles for Interstellar Absorption Lines}",
      journal = {\apj},
         year = 1991,
        month = sep,
       volume = {379},
        pages = {245},
          doi = {10.1086/170498},
       adsurl = {https://ui.adsabs.harvard.edu/abs/1991ApJ...379..245S}
}

@ARTICLE{lehner2020,
       author = {{Lehner}, Nicolas and {Berek}, Samantha C. and {Howk}, J. Christopher and {Wakker}, Bart P. and {Tumlinson}, Jason and {Jenkins}, Edward B. and {Prochaska}, J. Xavier and {Augustin}, Ramona and {Ji}, Suoqing and {Faucher-Gigu{\`e}re}, Claude-Andr{\'e} and {Hafen}, Zachary and {Peeples}, Molly S. and {Barger}, Kat A. and {Berg}, Michelle A. and {Bordoloi}, Rongmon and {Brown}, Thomas M. and {Fox}, Andrew J. and {Gilbert}, Karoline M. and {Guhathakurta}, Puragra and {Kalirai}, Jason S. and {Lockman}, Felix J. and {O'Meara}, John M. and {Pisano}, D.~J. and {Ribaudo}, Joseph and {Werk}, Jessica K.},
        title = "{Project AMIGA: The Circumgalactic Medium of Andromeda}",
      journal = {\apj},
         year = 2020,
        month = sep,
       volume = {900},
       number = {1},
          eid = {9},
        pages = {9},
          doi = {10.3847/1538-4357/aba49c},
archivePrefix = {arXiv},
       eprint = {2002.07818},
 primaryClass = {astro-ph.GA},
       adsurl = {https://ui.adsabs.harvard.edu/abs/2020ApJ...900....9L}
}

@ARTICLE{krishnarao2022,
       author = {{Krishnarao}, Dhanesh and {Fox}, Andrew J. and {D'Onghia}, Elena and {Wakker}, Bart P. and {Cashman}, Frances H. and {Howk}, J. Christopher and {Lucchini}, Scott and {French}, David M. and {Lehner}, Nicolas},
        title = "{Observations of a Magellanic Corona}",
      journal = {\nat},
         year = 2022,
        month = sep,
       volume = {609},
       number = {7929},
        pages = {915-918},
          doi = {10.1038/s41586-022-05090-5},
archivePrefix = {arXiv},
       eprint = {2209.15017},
 primaryClass = {astro-ph.GA},
       adsurl = {https://ui.adsabs.harvard.edu/abs/2022Natur.609..915K}
}

@ARTICLE{lehner2015,
       author = {{Lehner}, Nicolas and {Howk}, J. Christopher and {Wakker}, Bart P.},
        title = "{Evidence for a Massive, Extended Circumgalactic Medium Around the Andromeda Galaxy}",
      journal = {\apj},
         year = 2015,
        month = may,
       volume = {804},
       number = {2},
          eid = {79},
        pages = {79},
          doi = {10.1088/0004-637X/804/2/79},
archivePrefix = {arXiv},
       eprint = {1404.6540},
 primaryClass = {astro-ph.GA},
       adsurl = {https://ui.adsabs.harvard.edu/abs/2015ApJ...804...79L}
}

@ARTICLE{richter2017,
       author = {{Richter}, P. and {Nuza}, S.~E. and {Fox}, A.~J. and {Wakker}, B.~P. and {Lehner}, N. and {Ben Bekhti}, N. and {Fechner}, C. and {Wendt}, M. and {Howk}, J.~C. and {Muzahid}, S. and {Ganguly}, R. and {Charlton}, J.~C.},
        title = "{An HST/COS legacy survey of high-velocity ultraviolet absorption in the Milky Way's circumgalactic medium and the Local Group}",
      journal = {\aap},
         year = 2017,
        month = nov,
       volume = {607},
          eid = {A48},
        pages = {A48},
          doi = {10.1051/0004-6361/201630081},
archivePrefix = {arXiv},
       eprint = {1611.07024},
 primaryClass = {astro-ph.GA},
       adsurl = {https://ui.adsabs.harvard.edu/abs/2017A&A...607A..48R}
}

@ARTICLE{morton2003,
       author = {{Morton}, Donald C.},
        title = "{Atomic Data for Resonance Absorption Lines. III. Wavelengths Longward of the Lyman Limit for the Elements Hydrogen to Gallium}",
      journal = {\apjs},
         year = 2003,
        month = nov,
       volume = {149},
       number = {1},
        pages = {205-238},
          doi = {10.1086/377639},
       adsurl = {https://ui.adsabs.harvard.edu/abs/2003ApJS..149..205M}
}

@ARTICLE{cashman2017,
       author = {{Cashman}, Frances H. and {Kulkarni}, Varsha P. and {Kisielius}, Romas and {Ferland}, Gary J. and {Bogdanovich}, Pavel},
        title = "{Atomic Data Revisions for Transitions Relevant to Observations of Interstellar, Circumgalactic, and Intergalactic Matter}",
      journal = {\apjs},
         year = 2017,
        month = may,
       volume = {230},
       number = {1},
          eid = {8},
        pages = {8},
          doi = {10.3847/1538-4365/aa6d84},
archivePrefix = {arXiv},
       eprint = {1701.04847},
 primaryClass = {astro-ph.GA},
       adsurl = {https://ui.adsabs.harvard.edu/abs/2017ApJS..230....8C}
}

@ARTICLE{tumlinson2017,
       author = {{Tumlinson}, Jason and {Peeples}, Molly S. and {Werk}, Jessica K.},
        title = "{The Circumgalactic Medium}",
      journal = {\araa},
         year = 2017,
        month = aug,
       volume = {55},
       number = {1},
        pages = {389-432},
          doi = {10.1146/annurev-astro-091916-055240},
archivePrefix = {arXiv},
       eprint = {1709.09180},
 primaryClass = {astro-ph.GA},
       adsurl = {https://ui.adsabs.harvard.edu/abs/2017ARA&A..55..389T}
}

@ARTICLE{putman2012,
       author = {{Putman}, M.~E. and {Peek}, J.~E.~G. and {Joung}, M.~R.},
        title = "{Gaseous Galaxy Halos}",
      journal = {\araa},
         year = 2012,
        month = sep,
       volume = {50},
        pages = {491-529},
          doi = {10.1146/annurev-astro-081811-125612},
archivePrefix = {arXiv},
       eprint = {1207.4837},
 primaryClass = {astro-ph.GA},
       adsurl = {https://ui.adsabs.harvard.edu/abs/2012ARA&A..50..491P}
}

@ARTICLE{mishra2024a,
       author = {{Mishra}, Sapna and {Muzahid}, Sowgat and {Dutta}, Sayak and {Srianand}, Raghunathan and {Charlton}, Jane},
        title = "{Characterizing cool, neutral gas, and ionized metals in the outskirts of low-z galaxy clusters}",
      journal = {\mnras},
         year = 2024,
        month = jan,
       volume = {527},
       number = {2},
        pages = {3858-3875},
          doi = {10.1093/mnras/stad3454},
archivePrefix = {arXiv},
       eprint = {2305.05698},
 primaryClass = {astro-ph.GA},
       adsurl = {https://ui.adsabs.harvard.edu/abs/2024MNRAS.527.3858M}
}

@ARTICLE{fox2005,
       author = {{Fox}, Andrew J. and {Wakker}, Bart P. and {Savage}, Blair D. and {Tripp}, Todd M. and {Sembach}, Kenneth R. and {Bland-Hawthorn}, Joss},
        title = "{Multiphase High-Velocity Clouds toward HE 0226-4110 and PG 0953+414}",
      journal = {\apj},
         year = 2005,
        month = sep,
       volume = {630},
       number = {1},
        pages = {332-354},
          doi = {10.1086/431915},
archivePrefix = {arXiv},
       eprint = {astro-ph/0505299},
 primaryClass = {astro-ph},
       adsurl = {https://ui.adsabs.harvard.edu/abs/2005ApJ...630..332F}
}

@ARTICLE{ashley2024,
       author = {{Ashley}, Trisha and {Fox}, Andrew J. and {Lockman}, Felix J. and {Wakker}, Bart P. and {Richter}, Philipp and {French}, David M. and {Moss}, Vanessa A. and {McClure-Griffiths}, Naomi M.},
        title = "{The Metallicities of Five Small High-velocity Clouds}",
      journal = {\apj},
         year = 2024,
        month = jan,
       volume = {961},
       number = {1},
          eid = {94},
        pages = {94},
          doi = {10.3847/1538-4357/ad0cb7},
archivePrefix = {arXiv},
       eprint = {2311.09377},
 primaryClass = {astro-ph.GA},
       adsurl = {https://ui.adsabs.harvard.edu/abs/2024ApJ...961...94A}
}

@ARTICLE{foley2016,
       author = {{Foley}, A.~R. and {Alberts}, T. and {Armstrong}, R.~P. and {Barta}, A. and {Bauermeister}, E.~F. and {Bester}, H. and {Blose}, S. and {Booth}, R.~S. and {Botha}, D.~H. and {Buchner}, S.~J. and {Carignan}, C. and {Cheetham}, T. and {Cloete}, K. and {Coreejes}, G. and {Crida}, R.~C. and {Cross}, S.~D. and {Curtolo}, F. and {Dikgale}, A. and {de Villiers}, M.~S. and {du Toit}, L.~J. and {Esterhuyse}, S.~W.~P. and {Fanaroff}, B. and {Fender}, R.~P. and {Fijalkowski}, M. and {Fourie}, D. and {Frank}, B. and {George}, D. and {Gibbs}, P. and {Goedhart}, S. and {Grobbelaar}, J. and {Gumede}, S.~C. and {Herselman}, P. and {Hess}, K.~M. and {Hoek}, N. and {Horrell}, J. and {Jonas}, J.~L. and {Jordaan}, J.~D.~B. and {Julie}, R. and {Kapp}, F. and {Kotz{\'e}}, P. and {Kusel}, T. and {Langman}, A. and {Lehmensiek}, R. and {Liebenberg}, D. and {Liebenberg}, I.~J.~V. and {Loots}, A. and {Lord}, R.~T. and {Lucero}, D.~M. and {Ludick}, J. and {Macfarlane}, P. and {Madlavana}, M. and {Magnus}, L. and {Magozore}, C. and {Malan}, J.~A. and {Manley}, J.~R. and {Marais}, L. and {Marais}, N. and {Marais}, S.~J. and {Maree}, M. and {Martens}, A. and {Mokone}, O. and {Moss}, V. and {Mthembu}, S. and {New}, W. and {Nicholson}, G.~D. and {van Niekerk}, P.~C. and {Oozeer}, N. and {Passmoor}, S.~S. and {Peens-Hough}, A. and {Pi{\'n}ska}, A.~B. and {Prozesky}, P. and {Rajan}, S. and {Ratcliffe}, S. and {Renil}, R. and {Richter}, L.~L. and {Rosekrans}, D. and {Rust}, A. and {Schr{\"o}der}, A.~C. and {Schwardt}, L.~C. and {Seranyane}, S. and {Serylak}, M. and {Shepherd}, D.~S. and {Siebrits}, R. and {Sofeya}, L. and {Spann}, R. and {Springbok}, R. and {Swart}, P.~S. and {Thondikulam}, Venkatasubramani L. and {Theron}, I.~P. and {Tiplady}, A. and {Toruvanda}, O. and {Tshongweni}, S. and {van den Heever}, L. and {van der Merwe}, C. and {van Rooyen}, R. and {Wakhaba}, S. and {Walker}, A.~L. and {Welz}, M. and {Williams}, L. and {Wolleben}, M. and {Woudt}, P.~A. and {Young}, N.~J. and {Zwart}, J.~T.~L.},
        title = "{Engineering and science highlights of the KAT-7 radio telescope}",
      journal = {\mnras},
         year = 2016,
        month = aug,
       volume = {460},
       number = {2},
        pages = {1664-1679},
          doi = {10.1093/mnras/stw1040},
archivePrefix = {arXiv},
       eprint = {1606.02929},
 primaryClass = {astro-ph.IM},
       adsurl = {https://ui.adsabs.harvard.edu/abs/2016MNRAS.460.1664F}
}

@ARTICLE{lehner2025,
       author = {{Lehner}, Nicolas and {Howk}, J. Christopher and {Collins}, Lucy and {Sameer} and {Wakker}, Bart P. and {Augustin}, Ramona and {Barger}, Kathleen A. and {Berg}, Michelle A. and {Bordoloi}, Rongmon and {Brown}, Thomas M. and {Cashman}, Frances H. and {Faucher-Gigu{\`e}re}, Claude-Andr{\'e} and {Fox}, Andrew J. and {French}, David M. and {Gilbert}, Karoline M. and {Guhathakurta}, Puragra and {O'Meara}, John M. and {O'Shea}, Brian W. and {Peeples}, Molly S. and {Pisano}, D.~J. and {Prochaska}, J. Xavier and {Stern}, Jonathan and {Tumlinson}, Jason and {Werk}, Jessica K. and {Williams}, Benjamin F.},
        title = "{Project AMIGA: The Inner Circumgalactic Medium of Andromeda from Thick Disk to Halo}",
      journal = {arXiv e-prints},
         year = 2025,
        month = jun,
          eid = {arXiv:2506.16573},
        pages = {arXiv:2506.16573},
          doi = {10.48550/arXiv.2506.16573},
archivePrefix = {arXiv},
       eprint = {2506.16573},
 primaryClass = {astro-ph.GA},
       adsurl = {https://ui.adsabs.harvard.edu/abs/2025arXiv250616573L}
}

@ARTICLE{hunter2012,
       author = {{Hunter}, Deidre A. and {Ficut-Vicas}, Dana and {Ashley}, Trisha and {Brinks}, Elias and {Cigan}, Phil and {Elmegreen}, Bruce G. and {Heesen}, Volker and {Herrmann}, Kimberly A. and {Johnson}, Megan and {Oh}, Se-Heon and {Rupen}, Michael P. and {Schruba}, Andreas and {Simpson}, Caroline E. and {Walter}, Fabian and {Westpfahl}, David J. and {Young}, Lisa M. and {Zhang}, Hong-Xin},
        title = "{Little Things}",
      journal = {\aj},
         year = 2012,
        month = nov,
       volume = {144},
       number = {5},
          eid = {134},
        pages = {134},
          doi = {10.1088/0004-6256/144/5/134},
archivePrefix = {arXiv},
       eprint = {1208.5834},
 primaryClass = {astro-ph.GA},
       adsurl = {https://ui.adsabs.harvard.edu/abs/2012AJ....144..134H}
}

@ARTICLE{jenkins2009,
       author = {{Jenkins}, Edward B.},
        title = "{A Unified Representation of Gas-Phase Element Depletions in the Interstellar Medium}",
      journal = {\apj},
         year = 2009,
        month = aug,
       volume = {700},
       number = {2},
        pages = {1299-1348},
          doi = {10.1088/0004-637X/700/2/1299},
archivePrefix = {arXiv},
       eprint = {0905.3173},
 primaryClass = {astro-ph.GA},
       adsurl = {https://ui.adsabs.harvard.edu/abs/2009ApJ...700.1299J}
}

@ARTICLE{bq2013,
       author = {{For}, Bi-Qing and {Staveley-Smith}, Lister and {McClure-Griffiths}, N.~M.},
        title = "{Galactic All-Sky Survey High-velocity Clouds in the Region of the Magellanic Leading Arm}",
      journal = {\apj},
         year = 2013,
        month = feb,
       volume = {764},
       number = {1},
          eid = {74},
        pages = {74},
          doi = {10.1088/0004-637X/764/1/74},
archivePrefix = {arXiv},
       eprint = {1208.5583},
 primaryClass = {astro-ph.GA},
       adsurl = {https://ui.adsabs.harvard.edu/abs/2013ApJ...764...74F}
}

@ARTICLE{kim2024,
       author = {{Kim}, Doyeon A. and {Zheng}, Yong and {Putman}, Mary E.},
        title = "{Identifying H I Emission and UV Absorber Associations near the Magellanic Stream}",
      journal = {\apj},
         year = 2024,
        month = may,
       volume = {966},
       number = {1},
          eid = {134},
        pages = {134},
          doi = {10.3847/1538-4357/ad2def},
archivePrefix = {arXiv},
       eprint = {2402.08810},
 primaryClass = {astro-ph.GA},
       adsurl = {https://ui.adsabs.harvard.edu/abs/2024ApJ...966..134K}
}

@ARTICLE{1996ARA&A..34..279S,
       author = {{Savage}, Blair D. and {Sembach}, Kenneth R.},
        title = "{Interstellar Abundances from Absorption-Line Observations with the Hubble Space Telescope}",
      journal = {\araa},
         year = 1996,
        month = jan,
       volume = {34},
        pages = {279-330},
          doi = {10.1146/annurev.astro.34.1.279},
       adsurl = {https://ui.adsabs.harvard.edu/abs/1996ARA&A..34..279S}
}

@ARTICLE{qu2019,
       author = {{Qu}, Zhijie and {Bregman}, Joel N. and {Hodges-Kluck}, Edmund J.},
        title = "{HST/COS Observations of the Warm Ionized Gaseous Halo of NGC 891}",
      journal = {\apj},
         year = 2019,
        month = may,
       volume = {876},
       number = {2},
          eid = {101},
        pages = {101},
          doi = {10.3847/1538-4357/ab17df},
archivePrefix = {arXiv},
       eprint = {1904.04716},
 primaryClass = {astro-ph.GA},
       adsurl = {https://ui.adsabs.harvard.edu/abs/2019ApJ...876..101Q}
}

@ARTICLE{richter2016,
       author = {{Richter}, P. and {Wakker}, B.~P. and {Fechner}, C. and {Herenz}, P. and {Tepper-Garc{\'\i}a}, T. and {Fox}, A.~J.},
        title = "{An HST/COS legacy survey of intervening Si III absorption in the extended gaseous halos of low-redshift galaxies}",
      journal = {\aap},
         year = 2016,
        month = may,
       volume = {590},
          eid = {A68},
        pages = {A68},
          doi = {10.1051/0004-6361/201527038},
archivePrefix = {arXiv},
       eprint = {1507.06317},
 primaryClass = {astro-ph.CO},
       adsurl = {https://ui.adsabs.harvard.edu/abs/2016A&A...590A..68R}
}

@ARTICLE{piacitelli2025,
       author = {{Piacitelli}, Daniel R. and {Brooks}, Alyson M. and {Christensen}, Charlotte and {Sanchez}, N. Nicole and {Faerman}, Yakov and {Shen}, Sijing and {Cruz}, Akaxia and {Keller}, Ben and {Quinn}, Thomas R. and {Wadsley}, James},
        title = "{Marvelous Metals: Surveying the Circumgalactic Medium of Simulated Dwarf Galaxies}",
      journal = {\apj},
         year = 2025,
        month = nov,
       volume = {993},
       number = {2},
          eid = {230},
        pages = {230},
          doi = {10.3847/1538-4357/ae06a0},
archivePrefix = {arXiv},
       eprint = {2505.08861},
 primaryClass = {astro-ph.GA},
       adsurl = {https://ui.adsabs.harvard.edu/abs/2025ApJ...993..230P}
}

@ARTICLE{gerasimov2024,
       author = {{Gerasimov}, Ivan S. and {Egorov}, Oleg V. and {Moiseev}, Alexei V. and {Kniazev}, Alexei Yu and {Lozinskaya}, Tatiana A. and {Egorova}, Evgeniya S.},
        title = "{Stellar feedback impact on the ionized gas kinematics in the dwarf galaxy Sextans B}",
      journal = {\mnras},
         year = 2024,
        month = apr,
       volume = {529},
       number = {2},
        pages = {1138-1153},
          doi = {10.1093/mnras/stae462},
archivePrefix = {arXiv},
       eprint = {2402.03027},
 primaryClass = {astro-ph.GA},
       adsurl = {https://ui.adsabs.harvard.edu/abs/2024MNRAS.529.1138G}
}

@ARTICLE{hunter2010,
       author = {{Hunter}, Deidre A. and {Elmegreen}, Bruce G. and {Ludka}, Bonnie C.},
        title = "{Galex Ultraviolet Imaging of Dwarf Galaxies and Star Formation Rates}",
      journal = {\aj},
         year = 2010,
        month = feb,
       volume = {139},
       number = {2},
        pages = {447-475},
          doi = {10.1088/0004-6256/139/2/447},
archivePrefix = {arXiv},
       eprint = {0911.4319},
 primaryClass = {astro-ph.CO},
       adsurl = {https://ui.adsabs.harvard.edu/abs/2010AJ....139..447H}
}

@ARTICLE{stocke2013,
       author = {{Stocke}, John T. and {Keeney}, Brian A. and {Danforth}, Charles W. and {Shull}, J. Michael and {Froning}, Cynthia S. and {Green}, James C. and {Penton}, Steven V. and {Savage}, Blair D.},
        title = "{Characterizing the Circumgalactic Medium of Nearby Galaxies with HST/COS and HST/STIS Absorption-line Spectroscopy}",
      journal = {\apj},
         year = 2013,
        month = feb,
       volume = {763},
       number = {2},
          eid = {148},
        pages = {148},
          doi = {10.1088/0004-637X/763/2/148},
archivePrefix = {arXiv},
       eprint = {1212.5658},
 primaryClass = {astro-ph.CO},
       adsurl = {https://ui.adsabs.harvard.edu/abs/2013ApJ...763..148S}
}

@ARTICLE{sembach1992,
       author = {{Sembach}, Kenneth R. and {Savage}, Blair D.},
        title = "{Observations of Highly Ionized Gas in the Galactic Halo}",
      journal = {\apjs},
         year = 1992,
        month = nov,
       volume = {83},
        pages = {147},
          doi = {10.1086/191734},
       adsurl = {https://ui.adsabs.harvard.edu/abs/1992ApJS...83..147S}
}

@INCOLLECTION{cosdhb2022,
       author = {{Soderblom}, David},
        title = "{COS Data Handbook v. 5.1 (Baltimore: STScI)}",
    booktitle = {COS Data Handbook v. 5.1 (Baltimore: STScI)},
         year = 2022,
       volume = {5},
       adsurl = {https://ui.adsabs.harvard.edu/abs/2022cosd.book..5.1S}
}

@ARTICLE{johnson2026,
       author = {{Johnson}, Sean D. and {Mishra}, Nishant and {Muzahid}, Sowgat and {Rudie}, Gwen C. and {Zahedy}, Fakhri S. and {Qu}, Zhijie and {Faucher-Gigu{\`e}re}, Claude-Andr{\'e} and {Stern}, Jonathan and {Li}, Jennifer I-Hsiu and {Fuller}, Elise and {Cantalupo}, Sebastiano and {Chen}, Hsiao-Wen and {Kadri}, Ahmad and {Kumar}, Suyash and {Will Liu}, Zhuoqi and {Walth}, Gregory},
        title = "{MUSEQuBES: Physical conditions, origins, and multi-element abundances of the circumgalactic medium of an isolated, star-forming dwarf galaxy at z=0.57}",
      journal = {arXiv e-prints},
         year = 2025,
        month = oct,
          eid = {arXiv:2510.06310},
        pages = {arXiv:2510.06310},
          doi = {10.48550/arXiv.2510.06310},
archivePrefix = {arXiv},
       eprint = {2510.06310},
 primaryClass = {astro-ph.GA},
       adsurl = {https://ui.adsabs.harvard.edu/abs/2025arXiv251006310J}
}

@ARTICLE{zahedy2021,
       author = {{Zahedy}, Fakhri S. and {Chen}, Hsiao-Wen and {Cooper}, Thomas M. and {Boettcher}, Erin and {Johnson}, Sean D. and {Rudie}, Gwen C. and {Chen}, Mandy C. and {Cantalupo}, Sebastiano and {Cooksey}, Kathy L. and {Faucher-Gigu{\`e}re}, Claude-Andr{\'e} and {Greene}, Jenny E. and {Lopez}, Sebastian and {Mulchaey}, John S. and {Penton}, Steven V. and {Petitjean}, Patrick and {Putman}, Mary E. and {Rafelski}, Marc and {Rauch}, Michael and {Schaye}, Joop and {Simcoe}, Robert A. and {Walth}, Gregory L.},
        title = "{The Cosmic Ultraviolet Baryon Survey (CUBS) - III. Physical properties and elemental abundances of Lyman-limit systems at z < 1}",
      journal = {\mnras},
         year = 2021,
        month = sep,
       volume = {506},
       number = {1},
        pages = {877-902},
          doi = {10.1093/mnras/stab1661},
archivePrefix = {arXiv},
       eprint = {2106.04608},
 primaryClass = {astro-ph.GA},
       adsurl = {https://ui.adsabs.harvard.edu/abs/2021MNRAS.506..877Z}
}

@ARTICLE{johnson2017,
       author = {{Johnson}, Sean D. and {Chen}, Hsiao-Wen and {Mulchaey}, John S. and {Schaye}, Joop and {Straka}, Lorrie A.},
        title = "{The Extent of Chemically Enriched Gas around Star-forming Dwarf Galaxies}",
      journal = {\apjl},
         year = 2017,
        month = nov,
       volume = {850},
       number = {1},
          eid = {L10},
        pages = {L10},
          doi = {10.3847/2041-8213/aa9370},
archivePrefix = {arXiv},
       eprint = {1710.06441},
 primaryClass = {astro-ph.GA},
       adsurl = {https://ui.adsabs.harvard.edu/abs/2017ApJ...850L..10J}
}
\bibliographystyle{aasjournal}

\begin{deluxetable*}{lllll llll}[!ht]
\small
\caption{Sextans B CGM Parameters derived from Voigt Profile Fitting}
\label{tab:vpfit_param} 
\tablehead{Quasar         &  RA          &   DEC      &   $\rho$  &     Ion       &    $v_{\rm LSR}$        &      $b$                    &    log\,$N_{\rm comp}$          &   log\,$N_{\rm tot}$\tm{a}  \\  
& (deg)  &  (deg)  & (kpc)  &  &   (km s$^{-1}$) & (km s$^{-1}$)  & ($N$ in cm$^{-2}$)  & ($N$ in cm$^{-2}$)}
\startdata
SDSSJ100035.48+052428.5   &    150.15    &    5.41    &    4.1    &    \HI\ (EBHIS)&   300                &    ...                       &    ...                       & 19.46$\pm$0.25\tm{b}      \\ 
(=\jshort)                          &              &            &           &    \HI\ (LAB)  &   300                &    ...                       &    ...                       &  19.25$\pm$0.22      \\ 
 & & & & \HI\ (KAT-7) & 300 &  ... &  ... & 18.0$\pm$0.2\tm{c}\\ 
 & & & & \HI\ (VLA)   & 300 &  ... &  ... & 18.0$\pm$0.2\tm{c}\\ 
			              &              &            &           &    \OI\        &   ...                   &    ...                       &    ...                       &  $<$13.52\tm{d}                    \\
                          &              &            &           &    \CII\       &   293$\pm$37    &    14$\pm$3      &    13.53$\pm$0.08   &     13.53$\pm$0.08                          \\
                          &              &            &           &    \SiII\      &   275$\pm$6    &    17$\pm$9      &    12.26$\pm$0.12   &   12.87$\pm$0.05          \\
                          &              &            &           &    \SiII\      &   296$\pm$3     &    15$\pm$4      &    12.61$\pm$0.06   &                              \\
                          &              &            &           &    \SiII\      &   327$\pm$15    &    14$\pm$7      &    12.19$\pm$0.11   &                              \\
                          &              &            &           &    \SiIII\     &   270$\pm$6     &    17$\pm$4      &    12.69$\pm$0.08   &   $>$13.29$\pm$0.04\tm{e}         \\
                          &              &            &           &    \SiIII\     &   293$\pm$2    &    23$\pm$5      &    13.15$\pm$0.04   &                              \\
                          &              &            &           &    \SiIII\     &   327$\pm$15    &    18$\pm$40     &    11.58$\pm$0.66   &                              \\
                          &              &            &           &    \SiIV\      &   296$\pm$5     &    30$\pm$7      &    13.02$\pm$0.08   &   13.02$\pm$0.08          \\
\hline
SDSSJ095915.65+050355.1   &   149.82   &   5.07   &  8.0   &  \HI\ (EBHIS)&   292                 &  ...                       &   ...                    &   $<$18.79  \\ 
(=\jtshort)                          &            &          &        &  \HI\ (LAB) &   292                  &   ...                      &   ...                    &  19.14$\pm$0.19   \\ 
 & & & & \HI\ (KAT-7) & 292 &  ... &  ... & $\sim$16--17\tm{c}\\ 
			              &            &          &        &  \OI\       &   ...                     &   ...                      &   ...                    &  $<$13.62\tm{d}                    \\
                          &            &          &        &  \CII\      &   292$\pm$10     &   15$\pm$6     &   13.28$\pm$0.11   &  13.28$\pm$0.11      \\
                          &            &          &        &  \SiII\     &   ...                     &   ...                      &   ...                    &  $<$12.13                \\
                          &            &          &        &  \SiIII\    &   286$\pm$2     &   17$\pm$2     &   12.58$\pm$0.04   &  12.58$\pm$0.04         \\
                          &            &          &        &  \SiIV\     &   ...                     &   ...                      &   ...                    & $<$12.43                  \\
                           &           &           &       &  \CIV\      &   ...                     &   ...                      &   ...                    &  $<$13.15                 \\        
\enddata
\tn{a}{Total Sextans B column density, integrated over components if multiple components are present. Upper limits are 3$\sigma$.}
\tn{b}{The \HI\ column densities are integrated over the velocity range $258<v_{\rm LSR}<348$\kms. Four \HI\ measurements are given, from EBHIS (16.2\arcmin\ beam), LAB (30\arcmin\ beam), KAT-7 (4.4$\times$3.2\arcmin\ beam), and VLA (6\arcsec) data.}
\tn{c}{The \HI\ column toward \jshort\ is constrained by KAT-7 \citep{namumba2018} and VLA \citep{hunter2012} data. Both give log\,$N$(\HI)$\approx$18. For \jtshort, extrapolating the KAT-7 data to 8 kpc gives log\,$N$(\HI)$\sim$16--17.}
\tn{d}{The \OI\ limit is based on a night-only reduction of the COS data, to exclude geocoronal emission.}
\tn{e}{The total \SiIII\ column toward \jshort\ is a lower limit because of saturation in $\lambda$1206.}
\end{deluxetable*}

\begin{deluxetable}{lll}[!ht]
\small
\caption{Silicon Mass and Gas Mass in the Sextans B CGM}
\label{tab:mass} 
\tablehead{Radius  &   $M_{\rm Si}^{\rm CGM}$\tm{a} & $M_{\rm gas}^{\rm CGM}$\tm{b}\\
   (kpc)    &  (10$^{2}$\msun)       &   (10$^{7}$\msun)}
\startdata
0--4~~~~~~~~~~~~~~~~~~~~~~  & 4.2 ~~~~~~~~~~~~~~~~~~~~~ &    3.0  \\
4--8  &    1.3  &    0.9  \\
0--8  &    5.5  &    3.9  \\
\enddata
\tn{a}{CGM silicon mass traced by \SiII, \SiIII, and \SiIV\ calculated using equation (1). The mass is given in two individual radial bins, plus a total across both bins.}
\tn{b}{Cool CGM mass calculated using the silicon abundance [Si/H]=$-$1.7 derived in Section~\ref{subsec:cloudy}.} 
\end{deluxetable}


\end{document}